\documentclass[]{aastex631}

\begin{document}

\title{Probing Velocity Dispersion inside CMEs in Inner Corona: New Insights on CME Initiation}

\author[0000-0002-6553-3807]{Satabdwa Majumdar}
\affiliation{Aryabhatta Research Institute of Observational Sciences, Nainital, 263001, India}
\affiliation{Austrian Space Weather Office, Geosphere Austria, Graz, Austria}

\author[0000-0002-2914-2040]{Elke D' Huys}
\affiliation{Solar-Terrestrial Centre of Excellence - SIDC, Royal Observatory of Belgium, Avenue Circulaire 3, Brussels 1180, Belgium}


\author[0000-0003-4105-7364]{Marilena Mierla}
\affiliation{Solar-Terrestrial Centre of Excellence - SIDC, Royal Observatory of Belgium, Avenue Circulaire 3, Brussels 1180, Belgium}
\affiliation{Institute of Geodynamics of the Romanian Academy, Bucharest, Romania}

\author[0000-0001-5859-5957]{Nitin Vashishtha}
\affiliation{Aryabhatta Research Institute of Observational Sciences, Nainital, 263001, India}

\author[0000-0002-9311-9021]{Dana-Camelia Talpeanu}
\affiliation{Solar-Terrestrial Centre of Excellence - SIDC, Royal Observatory of Belgium, Avenue Circulaire 3, Brussels 1180, Belgium}

\author[0000-0003-4653-6823]{Dipankar Banerjee}
\affiliation{Aryabhatta Research Institute of Observational Sciences, Nainital, 263001, India}
\affiliation{Indian Institute of Astrophysics,2nd Block, Koramangala, Bangalore, 560034, India}
\affiliation{Center of Excellence in Space Science, IISER Kolkata, Kolkata 741246, India}

\author[0000-0002-6362-5054]{Martin A. Reiss}
\affiliation{Community Coordinated Modeling Center, NASA Goddard Space Flight Center, 8800 Greenbelt Rd., Greenbelt, MD 20771, USA}



\begin{abstract}
This work studies the kinematics of the leading edge and the core of 6 Coronal Mass Ejections (CMEs) in the combined field of view of Sun Watcher using Active Pixel System detector and Image Processing (SWAP) on-board PRoject for On-Board Autonomy (PROBA-2) and the ground-based K-Cor coronagraph of the Mauna Loa Solar Observatory (MLSO). We report, for the first time, on the existence of a critical height h$_\mathrm{c}$, which marks the onset of velocity dispersion inside the CME. This height for the studied events lies between 1.4-1.8 R$_{\odot}$, in the inner corona. We find the critical heights to be relatively higher for gradual CMEs, as compared to impulsive ones, indicating that the early initiation of these two classes might be different physically. We find several interesting imprints of the velocity dispersion on CME kinematics. The critical height is strongly correlated with the flux-rope minor radius and the mass of the CME. Also, the magnitude of velocity dispersion shows a reasonable positive correlation with the above two parameters. We believe these results will advance our understanding of CME initiation mechanisms and will help provide improved constraints to CME initiation models.
\end{abstract}
\keywords{}


\section{Introduction} \label{intro}

The discovery of Coronal Mass Ejections (or CMEs) dates back to the early 1970s \citep{hansen_1971}, which marks a major milestone in the context of the Sun-Earth connection. Since their discovery \citep[for a historical review, see][]{gopalswamy_2016}, these violent, yet fascinating eruptions of plasma and magnetic field attracted the attention of the scientific community. This fascination to study CMEs arises from a two fold perspective. The first comes from scientific curiosity due to the wide range of properties exhibited by these eruptions \citep{webb_2012,chen_2017}, and the second arises from an economical and technological perspective. These eruptions are the major drivers of space weather, having the potential to create strong geomagnetic storms that can disrupt communication systems and pose threats to astronauts in space \cite[for a recent review, see][]{temmer_2021}, and thus a clear understanding of them is essential. \\

From describing CMEs as ``discrete, bright, white-light features" \citep[as quoted in][]{hundhausen_1984}, we have come a long way towards having a better and improved understanding of these eruptions. This has been possible thanks to the wealth of observational resources and models that have helped us probe into the different aspects of CMEs. Often observable properties of CMEs like height, width, etc. are estimated based on their morphological properties, and this morphological classification of CMEs is mostly done based on their appearance in the white-light coronagraph images \citep[see][]{munro_1979,Cremades_2004}. In this context, two particular morphologies stand out from the rest: the loop-like CME \citep{crifo_1983} and the three-part structured CME \citep{illing_1985}. The latter is considered evidence for a flux-rope structure of a CME \citep{song_2022}, and such a three part structure can be statistically found in one out of every three CMEs \citep{vourlidas_2013}.\\

Ever since the start of CME observations, understanding the kinematics of these eruptions has been of significant interest and importance. CMEs often show three phases of evolution in their kinematic profile, with an initial slow rise phase, followed by an impulsive acceleration phase and then the propagation phase where it experiences very little or almost no acceleration or deceleration \citep{Zhang_2001}. Out of these three phases, the first two phases are usually confined in the inner coronal heights ($\sim3$R$_{\odot}$) where the CME generally experiences the main acceleration phase \citep{chen_2003,majumdar_2020,majumdar_2022}. 
However, we are yet to arrive at a clear understanding of the first two phases of evolution, and these initial phases are important as they hold clues about the eruption process of the CME.\\

Since the three-part structured CMEs provide a unified morphological picture in the form of a flux-rope, it is important to track the different parts of these three-part CMEs. Such an attempt will shed more light on the relative kinematics of these different parts, and the role played by one on the evolution of the other (if any at all). Unfortunately, such studies \citep[see; ][]{schmahl_1977,wood_1999,nandita_2000,krall_2001,maricic_2004,koutchmy_2008,maricic_2009,chifu_2012} are rare, as it is extremely difficult to identify such structured CMEs in the lower heights. Nevertheless, these studies are crucial as they can provide exceptionally important views as inputs to different CME initiation models \citep[for a review, see][]{forbes_2006}. 
These studies have shown that the CME leading edge (hereafter LE) travels with a higher velocity than the CME core, giving rise to a velocity dispersion inside a CME. Particularly, \cite{maricic_2009} reported on the different times of onset of acceleration in the LE and the core. This indicates the possibility of the existence of a time and height at which the dispersion in velocities of the two structures starts. Also, recently, \cite{bemporad_2018} showed that a radial gradient of speed from the inner core to the LE can get manifested into a radial gradient of electron temperature distribution. This difference in temperature across different fronts might manifest into different expansion rates along the different parts of a CME, as recently reported by \cite{majumdar_2022}. Hence, even from the thermodynamic perspective, this dispersion of velocity across the core and the LE seems to play an important role. Thus, it is extremely important to study this velocity gradient within the CME, particularly in the lesser explored inner coronal heights, to better understand its effect on the different aspects of CME early evolution.\\

CMEs can be grouped into two dynamical classes, impulsive and gradual CMEs \citep{moon_2002}. The impulsive CMEs tend to originate from the active regions on the Sun, while the gradual CMEs are connected to erupting prominences. This distinction is reflected in different aspects of the kinematics of these CMEs \citep[see][]{pant_2021,majumdar_2021} and in various statistical properties as recently pointed out based on a CME source region catalogue by \cite{majumdar_2023}. Hence, whether any difference in the velocity dispersion is seen for impulsive and gradual CMEs is worth studying. In this work, we aim to probe this velocity dispersion of CMEs in the inner corona, and the influence of it on the overall kinematic evolution of CMEs. In Section~\ref{data}, we describe the data sources and the working method, followed by the main results of this work in Section~\ref{results}. We finally summarize the main conclusions in Section~\ref{conclusions}.

\section{Data and Method} \label{data}
\subsection{Data Source}
To track the CMEs in the inner corona, we use the combined observations in the 17.4 nm bandpass of the Sun Watcher using Active Pixel system detector and image processing \citep[SWAP;][]{halain_2013,seaton_2013} on-board PRoject for On-Board Autonomy \citep[PROBA-2;][]{santandrea_2013} and the 2 minute cadence level 2 data processed with Normalized Radially Graded Filter \citep[NRGF;][]{morgan_2006} from the ground based coronagraph K-Cor of the Mauna Loa Solar Observatory. We track the CME further out in the corona using the data from Large Angle Spectroscopic COronagraph C2 \citep[LASCO;][]{Brueckner_1995} on-board the Solar and Heliospheric Observatory \citep[SOHO;][]{domingo_1995}. We also identify the source regions of the CMEs using the different passbands of the Atmospheric Imaging Assembly \citep[AIA;][]{aia} on-board the Solar Dynamics Observatory \citep[SDO;][]{pesnell_2012} and Extreme Ultraviolet Imager \citep[EUVI;][]{howard_2008} on-board the Solar Terrestrial Relations Observatory \citep[STEREO;][]{stereo}. 

\subsection{Data Preparation}

\begin{figure*}[ht!]
\gridline{\fig{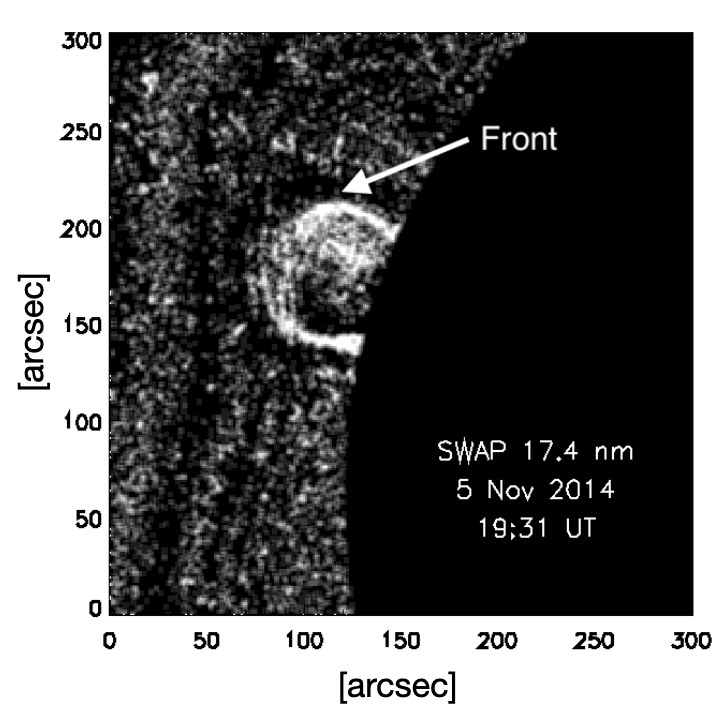}{0.44\textwidth}{(a)}
          \fig{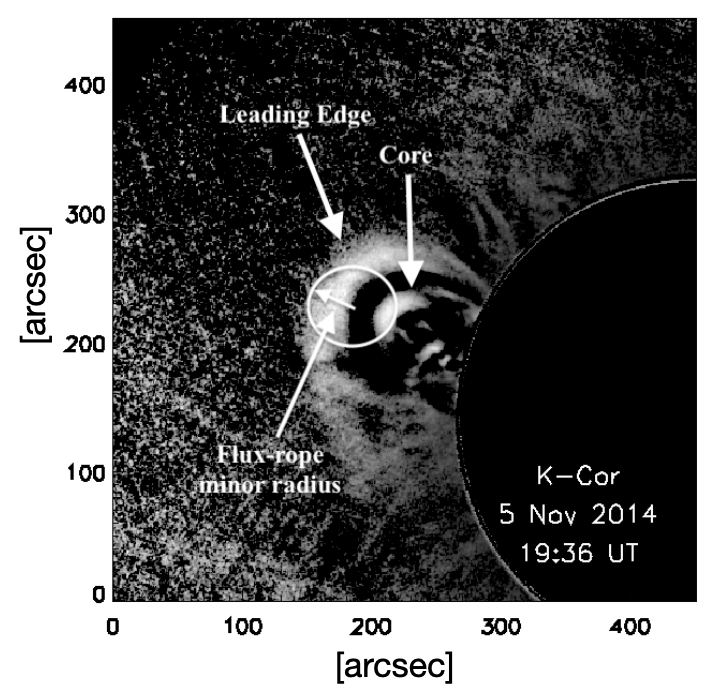}{0.44\textwidth}{(b)}
          }
\gridline{\fig{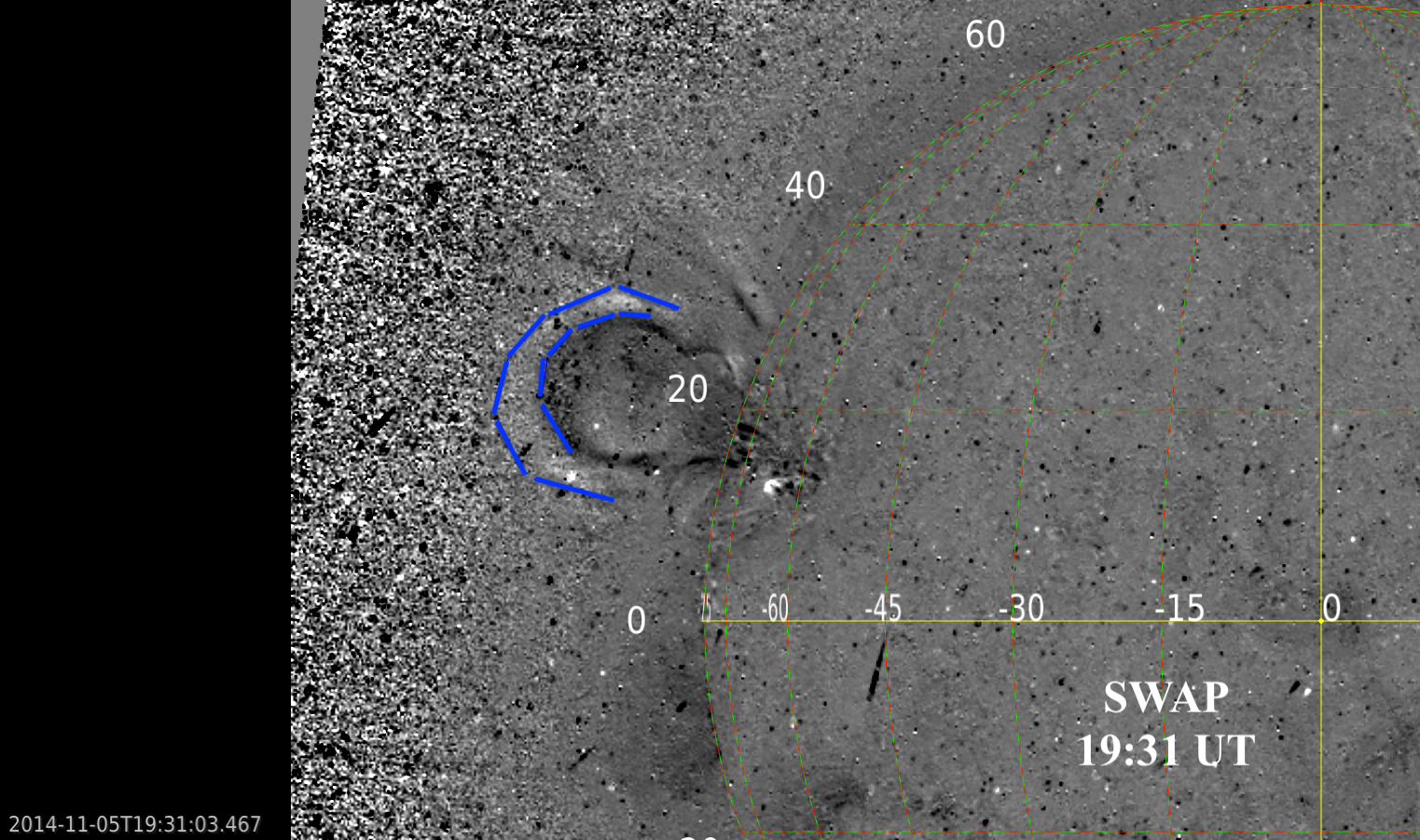}{0.44\textwidth}{(c)}
          \fig{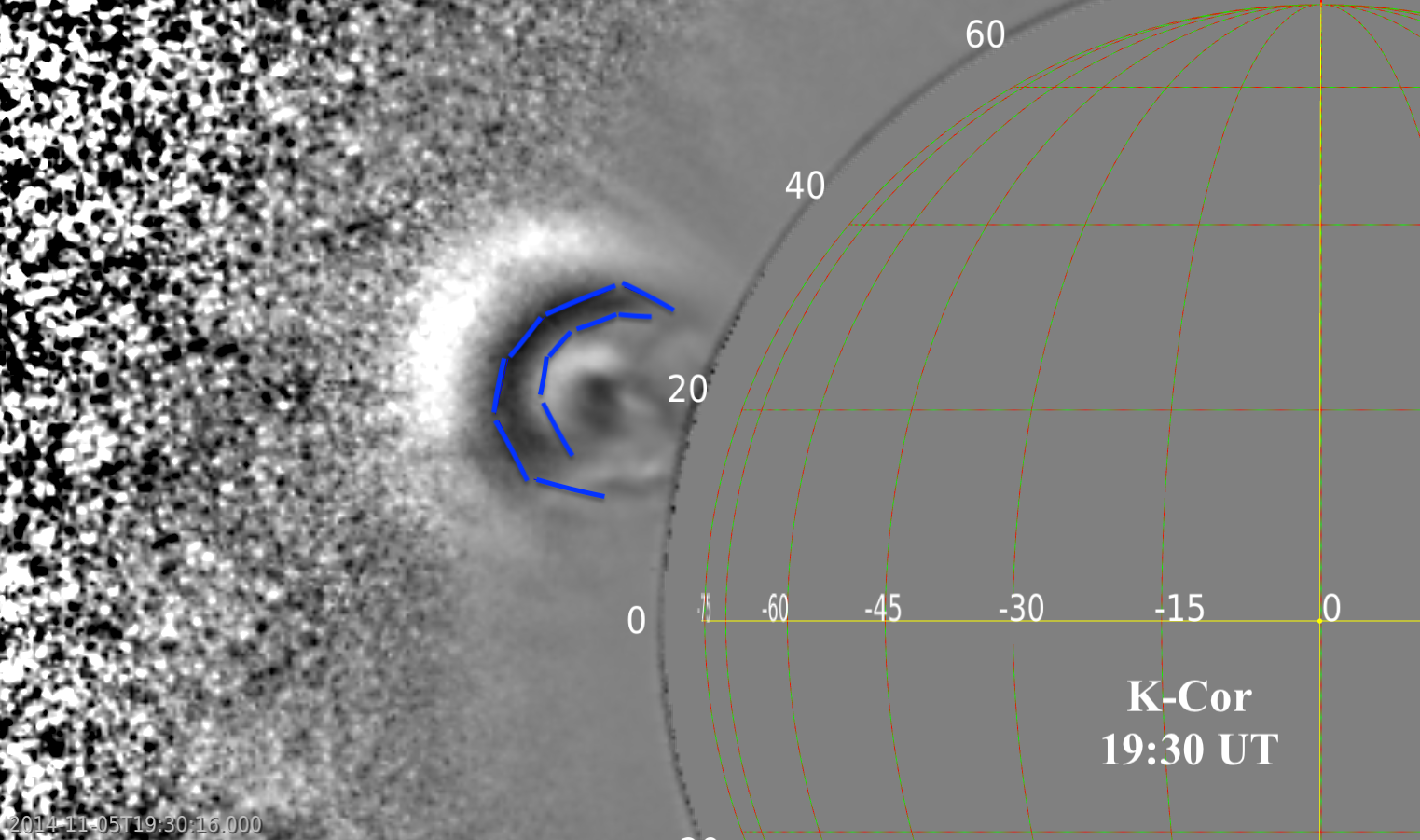}{0.44\textwidth}{(d)}
          }
\caption{A view of the eruption front for the CME that occurred on 5 November, 2014 in (a) SWAP FOV (up to 1.7 R$_{\odot}$ and in (b) the three-part structured CME in K-Cor FOV (1.05 - 3 R$_{\odot}$). A circle fitted to the cross-section of the CME is shown in (b) with the corresponding minor radius. (c) Running difference image in SWAP FOV (created using JHelioviewer) with blue lines tracing the eruption front, and (d) running difference image in K-Cor FOV (created using JHelioviewer) with Blue lines corresponding to (c) overlaid on the K-Cor FOV.
\label{cme}}
\end{figure*}

To increase the visibility of the off-limb structures in SWAP images, the intensity is re-scaled using the standard BYTSCL.PRO routine in IDL, which simply scales the values of an array lying in the range (min,max) to (0,new$\_$max), and then the disk of the Sun was masked. Running difference images are then created for the purpose of removing the static features. In order to further enhance the coronal intensity for outward propagating structures, a radial filter, called the Multi-Scale Gaussian Normalization \citep[MGN;][]{morgan_2014} has been applied. A resulting image is shown in Figure~\ref{cme}(a), where the eruption front for the CME on 5 November 2014, can be identified in the SWAP field of view (FOV). We also apply a gamma transformation in the form of $I^{\gamma}$ (where $I$ is the image array and $\gamma$ lies between 0.3 to 0.5) to the above images, to further enhance the brightness and the contrast of the eruption features with respect to the background. To remove the static features in the off-limb corona in K-Cor images, running difference images are created, where each image is subtracted by the previous image and so on. A resulting image is shown in Figure~\ref{cme}(b), where a three-part structure CME can be seen. To understand the spatial overlap of the CME as seen in SWAP and K-Cor, in Figure~\ref{cme}(c), a running difference image in SWAP FOV (created using JHelioviewer \citep{muller_2017}) is shown, with blue lines tracing the eruption front.  The same blue lines are then overlaid on the CME observed in K-Cor (running difference images created using JHelioviewer) in Figure~\ref{cme}(d), showing that SWAP approximately observes the core of the CME in K-Cor. It is worth noting that the blue outline lies slightly ahead of the core in K-Cor. This is partly due to the time difference between the two images and the fact that Extreme Ultraviolet and white light images might not be looking at the exact same physical feature, which is also seen as an outcome in the form of the height difference between the SWAP front and the K-Cor core as shown in Figure~\ref{h_diff}. However, if the K-Cor core height measured at 19:30 UT (which was 1.33 R$_{\odot}$), is translated forward in time with the instantaneous speed of the core at that time, taken from Figure~\ref{vel_prof}(a) (which was 320 km/s), the estimated height comes to be 1.36 R$_{\odot}$, which leads to a higher overlap with the overlaid blue contours. 


\begin{center}
\begin{table}[]
    \centering
\begin{tabular}{c|c|c|c|c|c|c}
   Date & Time (UT) & Latitude ($^{\circ}$) & Longitude ($^{\circ}$)  & Source Region & h$_c$ (R$_{\odot}$) & V$_{disp}$ (km/s)  \\
   \hline \hline
   2014 - 11 - 05 & 19:30:16 & 16 & -80 & AR & 1.54 & 115 \\
   2014 - 12 - 21 & 01:57:58 & 10 & -82 & AP & 1.42 & 152 \\
   2020 - 11 - 26 & 20:50:56 & 32 & -85 & AR & 1.59 & 140 \\
   2021 - 05 - 07 & 19:02:48 & 19 & -83 & AR & 1.55 & 166 \\
   2021 - 06 - 10 & 18:07:54 & 22 & -135 & AP & 1.82 & 217 \\
   2022 - 05 - 24 & 22:41:48 & 35 & -77 & PE & 1.7   & 159 \\
    \hline
\end{tabular} 
\caption{The table shows the summary of the studied events. The date and time of every event (derived from K-Cor images) is provided, followed by the latitude and longitude (in Stonyhurst heliographic coordinates) of the source region of the CME. The source region column (AR - Active Regions, AP - Active Prominences and PE - Prominence Eruptions) provides the class of the identified source region. The last two columns provide the critical height and the strength of velocity dispersion.}\label{tab1}
\end{table}
\end{center}


\subsection{Working Method}
For this study, only CMEs that showed clear three part structure in K-Cor and LASCO images were selected, which strongly restricted the event selection. It was further ensured that the CME was observed sideways in the plane of the sky, with a distinct leading edge that remained un-diluted in successive frames. K-Cor being a ground-based facility, data acquisition is restricted to only a few hours during the day, which further constrained the event selection. 
To identify the source regions of the CMEs, we follow the methodology of \cite{pant_2021} and \cite{majumdar_2023} on the SWAP and AIA images, and the identified source regions are shown in Table~\ref{tab1}. The Active Regions (AR) are the usual flaring regions showing bright emission in extreme ultraviolet wavelengths, the Prominences that erupt from the quiet Sun regions are denoted as PEs and the prominences whose foot-points are rooted in an Active Region are classified as APs. It can be seen from the values of the longitudes that almost all the CMEs studied in this work are limb events, (which is also an outcome of our event selection, as we selected events that show the three-part structure) except the one that occurred on June 10, 2021. 

\section{Results} \label{results}
\subsection{Observations}
\begin{figure*}[ht!]
\gridline{\fig{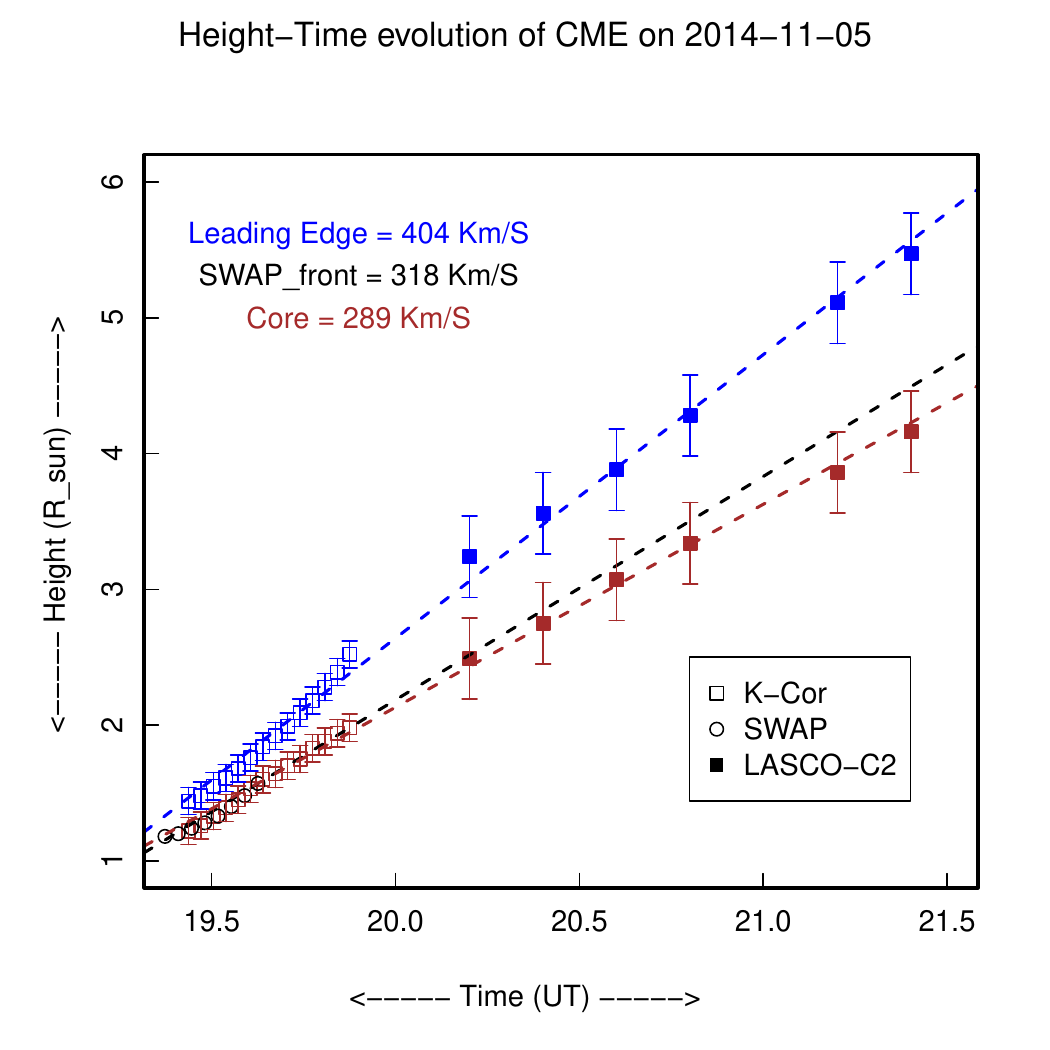}{0.48\textwidth}{(a)}
          \fig{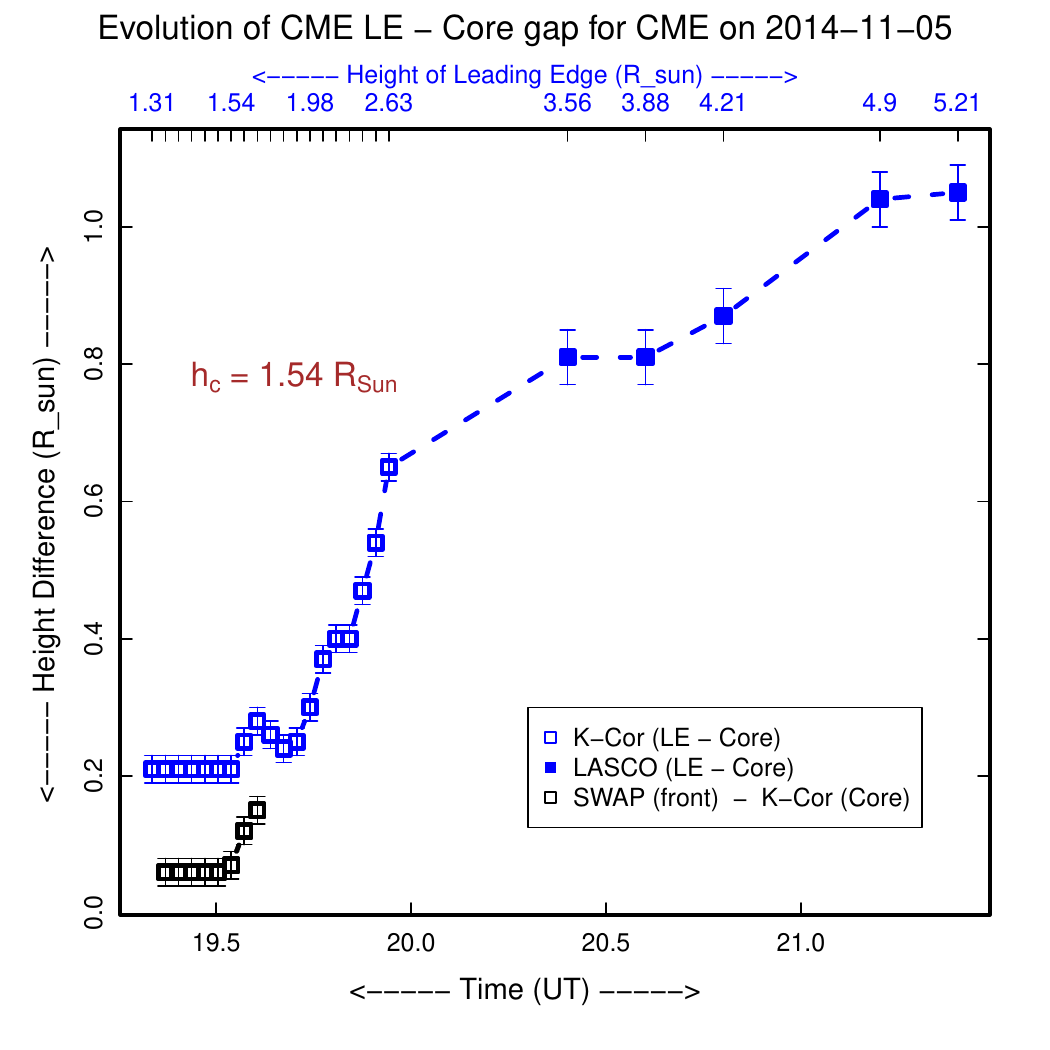}{0.48\textwidth}{(b)}
          }
\caption{(a) Height-time plot of the LE and core in blue and brown data points, for the CME on November 5, 2014 as tracked in the K-Cor FOV, and in black data points, the eruption front as tracked in the SWAP FOV. (b) Plot of the difference in height of LE and core for the same CME in the K-Cor FOV and LASCO FOV (in blue), and the difference in height of the SWAP front and K-Cor core (in black). The corresponding height of the LE is plotted on the top axis.}
\label{h_diff}
\end{figure*}

We manually tracked the LE and the core by eye, and in Figure~\ref{h_diff}(a), we plotted the height of the LE (in blue data points) and core (in brown data points) of the CME on November 5, 2014 as seen in K-Cor and LASCO-C2 FOV and the height of the eruption front in SWAP FOV (in black data points). We find that the height of the eruption front in SWAP FOV is very similar to the core height in K-Cor, thus confirming that SWAP is indeed observing the core of the CME (which was also noted in \cite{jennifer_2019}). 
From Figure~\ref{h_diff}(a), a curvature is clearly seen in the data points corresponding to K-Cor and SWAP observations, thus indicating the presence of acceleration at these heights. However, for an initial rough estimate of the linear speeds of the three different parts of the CME, we fit a straight line to the different height-time data and get the linear speeds of the K-Cor LE as 404 km s$^{-1}$, SWAP front as 318 km s$^{-1}$ and K-Cor core as 289 km s$^{-1}$. Thus, we see a radial gradient in velocity inside a CME from core to the LE, and hence a presence of velocity dispersion. It is indeed interesting to see that the speed of the eruption front in SWAP (that approximately corresponds to the core in K-Cor) is higher than the speed of the core in K-Cor. We believe this is an outcome of the fact that the extreme ultraviolet and white-light images do not necessarily capture the same physical feature, and as the SWAP front is tracked at the outer edge of the SWAP field of view, lack of proper signal to noise ratio in those regions would also contribute to the errors in height measurements. 
A more detailed investigation of this behavior in a future study in this regard, with higher resolution data that would enable tracking multiple structures inside a three part structured CME, and with better signal to noise ratio, would be crucial and helpful. In Table~\ref{tab1}, a measure of the magnitude of dispersion ($\mathrm{V_{disp}}$) is provided as the difference between the linear velocities of the LE and core in the K-Cor and LASCO FOV as shown in Figure~\ref{h_diff}(a). We do not include the velocity in the SWAP FOV, since SWAP does not observe the three-part structure of the CME. To have a better understanding of this dispersion in velocities, we plot the difference between the heights of the K-Cor LE and K-Cor core (in blue) and between SWAP front and K-Cor core (in black) versus time in Figure~\ref{h_diff}(b), with the corresponding height of the K-Cor LE on the top axis. 
It can be seen that initially, the gap between the tracked features remains constant, thus indicating that velocity dispersion has not yet started. We find that this regime continues until a critical height (h$_{c}$) of 1.54 R$_{\odot}$, corresponding to an approximate duration of 14 minutes since their first appearance in the SWAP FOV, after which, the gap starts increasing. This is the height of commencement of velocity dispersion . 
Thus it seems that as the flux-rope starts evolving, initially the entire structure moves with the same speed. It is only after a finite time that the change in speeds sets in, which marks the start of velocity dispersion. For the other five CMEs studied in this work, further examples of the snapshots of the CMEs as seen in the K-Cor and SWAP FOV, are included in the appendices, in Figure~\ref{append_fig1} and \ref{append_fig2}, and a plot of the difference in the LE and core heights for each of those CMEs is shown in Figure~\ref{append_fig4}. 

\begin{figure*}[ht!]
\gridline{
          \fig{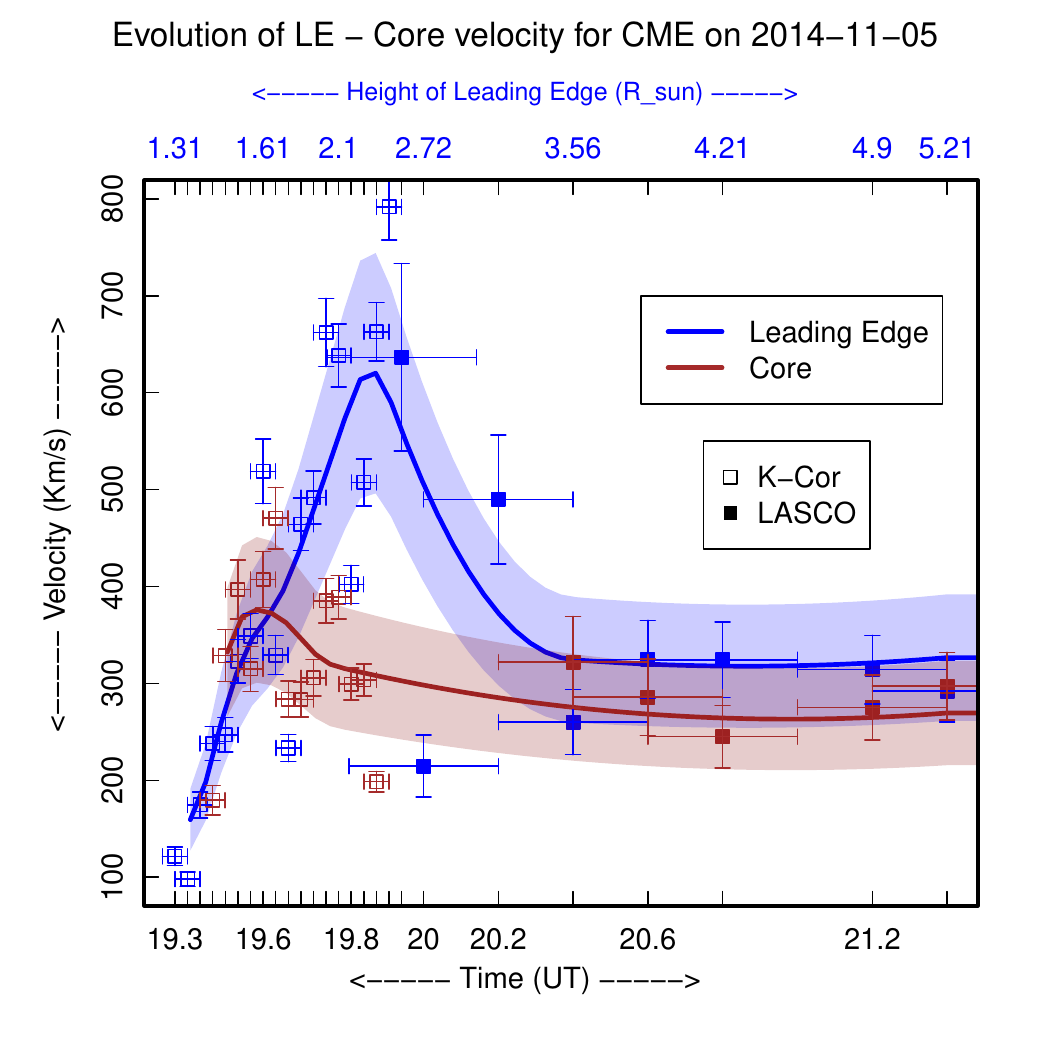}{0.48\textwidth}{(a)}
          \fig{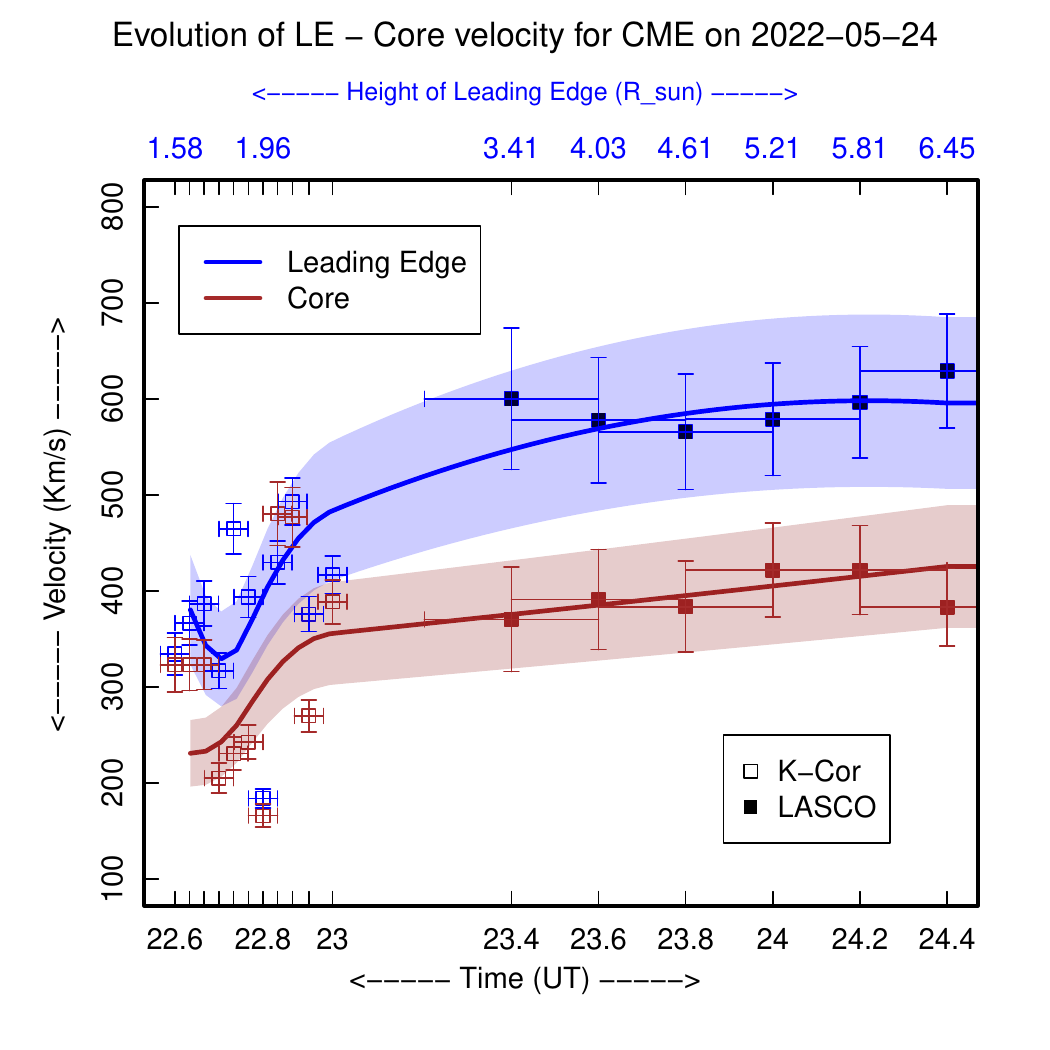}{0.48\textwidth}{(b)}
          }
\caption{A plot of the velocity profile of the LE and core in the combined K-Cor and LASCO FOV for the CME on November 5, 2014 in (a) and for the CME on May 24, 2022 in (b). The solid lines represent a smooth spline fitted to the data points, and the shaded region indicate the uncertainty in the fitted spline.
\label{vel_prof}}
\end{figure*}

In the context of velocity dispersion and the presence of a critical height, instead of relying on the linear speeds, it is also important to look into the velocity profiles of the two tracked features in K-Cor and LASCO FOV. The instantaneous velocity data points and profiles are calculated using smooth splines \citep[for details, see][]{majumdar_2020}. So, in Figure~\ref{vel_prof}(a) and (b), we plot the velocity profiles of the LE and the core of two CMEs (in blue and brown data points and lines respectively). The shaded region denote the uncertainty in the smooth spline fitting. In Figure~\ref{vel_prof}(a) for the CME on 5 November 2014, a distinct difference is noted in the two velocity profiles. We find that the LE velocity profile initially rises and then decreases, showing the presence of the typical impulsive acceleration phase that is known to occur in the inner corona \citep{majumdar_2020,majumdar_2022}. On the other hand, the velocity profile of the CME core shows similar behaviour but on a smaller scale of magnitude. This indicates the role of the injected Lorentz force in accelerating impulsive CMEs suddenly to very high speeds, while the core appears to experience a weaker acceleration, leading to a much lesser speed, thereby initiating the dispersion in velocity. However, when analyzing the velocity profiles of the LE and core for a different CME (24 May 2022) in Figure~\ref{vel_prof}(b), a different scenario is observed. The LE does not show any signs of impulsiveness, but a slow yet steady rise in the velocity is noted, which is a signature of gradual CMEs that experience relatively weaker acceleration but have a longer acceleration duration. The core, in this case, shows resemblance with the LE, as a steady increase in the core speed is also noted, although the speeds of the core and the LE are different. A plot of the velocity profiles of the LE and core for the other CMEs studied in this paper is included in the appendices, please see Figure~\ref{append_fig3}.\\
We find that the critical heights (i.e. the height at which the velocity dispersion starts) of all the investigated CMEs lie in the range of $1.42-1.82$ R$_{\odot}$ (see Table~\ref{tab1}), which is in the inner corona. We also note that CMEs originating from erupting prominences (PE or AP), tend to have higher critical heights as compared to CMEs originating from ARs, however, since this conclusion is based on very few events, in future it would be important to test this on a larger data set.\\

\subsection{Simulation}
\begin{figure*}[ht!]
\gridline{\fig{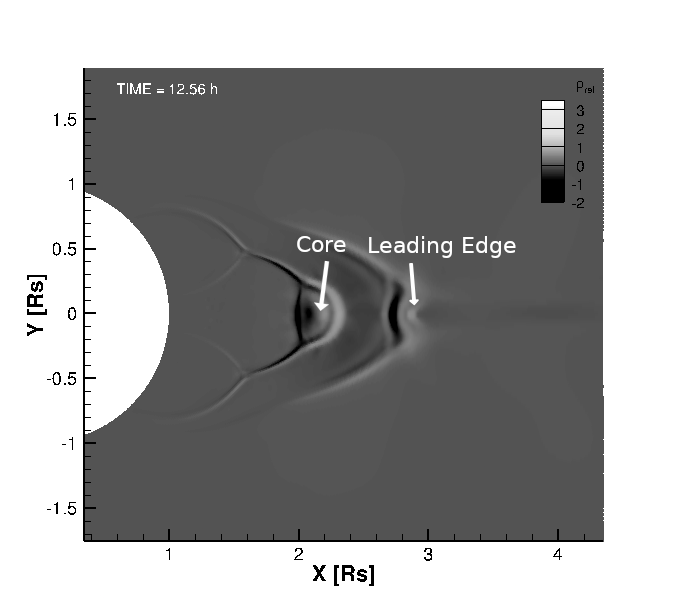}{0.48\textwidth}{(a)}                 \fig{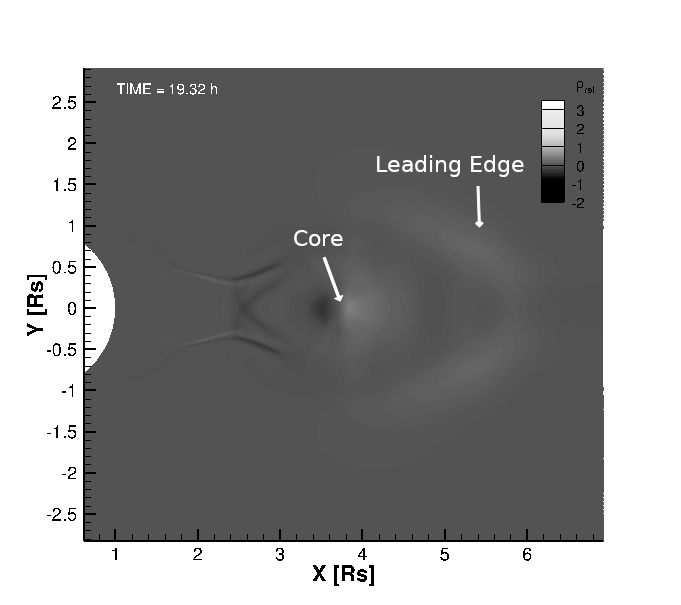}{0.48\textwidth}{(b)}}
\gridline{          \fig{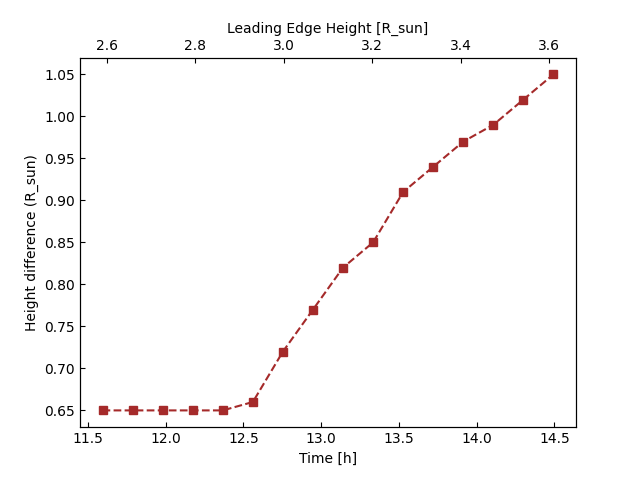}{0.48\textwidth}{(c)}
          }
\caption{(a,b) A simulated breakout CME for two different time frames, with the two arrows denoting the leading edge and core of the flux-rope. The referenced time is measured from the initiation of shearing. The color bar signifies the relative running difference density, denoted as $\rho_{\text{rel}} = \frac{\rho(t_i) - \rho(t_{i-1})}{\rho_0}$, where $\rho_0$ represents the density of the equilibrium state before applying the shear, and $\rho(t_i)$ represents the density in the $i$th time frame. (c) A plot of the difference in the height of the leading edge and flux-rope centre for the breakout CME simulated in (a,b) versus time, with the corresponding leading edge height plotted along the top axis.
\label{cartoon}}
\end{figure*}

We also conducted a 2.5D numerical simulation of a breakout CME \citep{van_der_Holst_2007,Zuccarello_2012,Hosteaux,dana2020,dana2022} using Message Passing Interface - Adaptive Mesh Refinement Versatile Advection Code (MPI-AMRVAC)  \citep{porth_o,Xia_2018} under a magnetohydrodynamic (MHD) regime. The initiation of the CME in the breakout scenario involves reconnection between the multipolar (quadrupolar) magnetic field and the overlying background magnetic field \citep{Antiochos_1998,Antiochos_1999}. To trigger the eruption, we induced shearing motions at the base of the central arcade, resulting in an upward movement of the arcade and reconnection with the overlying magnetic field. We tracked the leading edge and the core of the flux rope (shown with two arrows in Figure~\ref{cartoon}(a,b) for two different time frames), and the evolution of the gap between these two features is plotted in Figure~\ref{cartoon}(c). For the ease of visualization, the Color-map in Figure~\ref{cartoon}(a,b) represents the relative running difference density given by $\rho_{\text{rel}} = \frac{\rho(t_i) - \rho(t_{i-1})}{\rho_0}$,where the running difference frames in density, created by subtracting two successive frames at times $t_i$ and $t_{i-1}$, is divided by $\rho_0$, which represents the density of the equilibrium state before applying the shear. We again find that the gap between the two features remains constant initially, and a critical height of start of velocity dispersion is thereby noted at a height of 2.84 R$_\odot$, which lies in the inner corona (as found for the six CMEs from the observations). The average speed of the LE of the simulated CME was determined to be $\sim$ 200 km/s, which is also consistent with the earlier discussion that gradual CMEs tend to exhibit higher critical heights compared to impulsive CMEs.



\subsection{Dependence on Flux-rope Cross-section}

\begin{figure*}[ht!]
\gridline{\fig{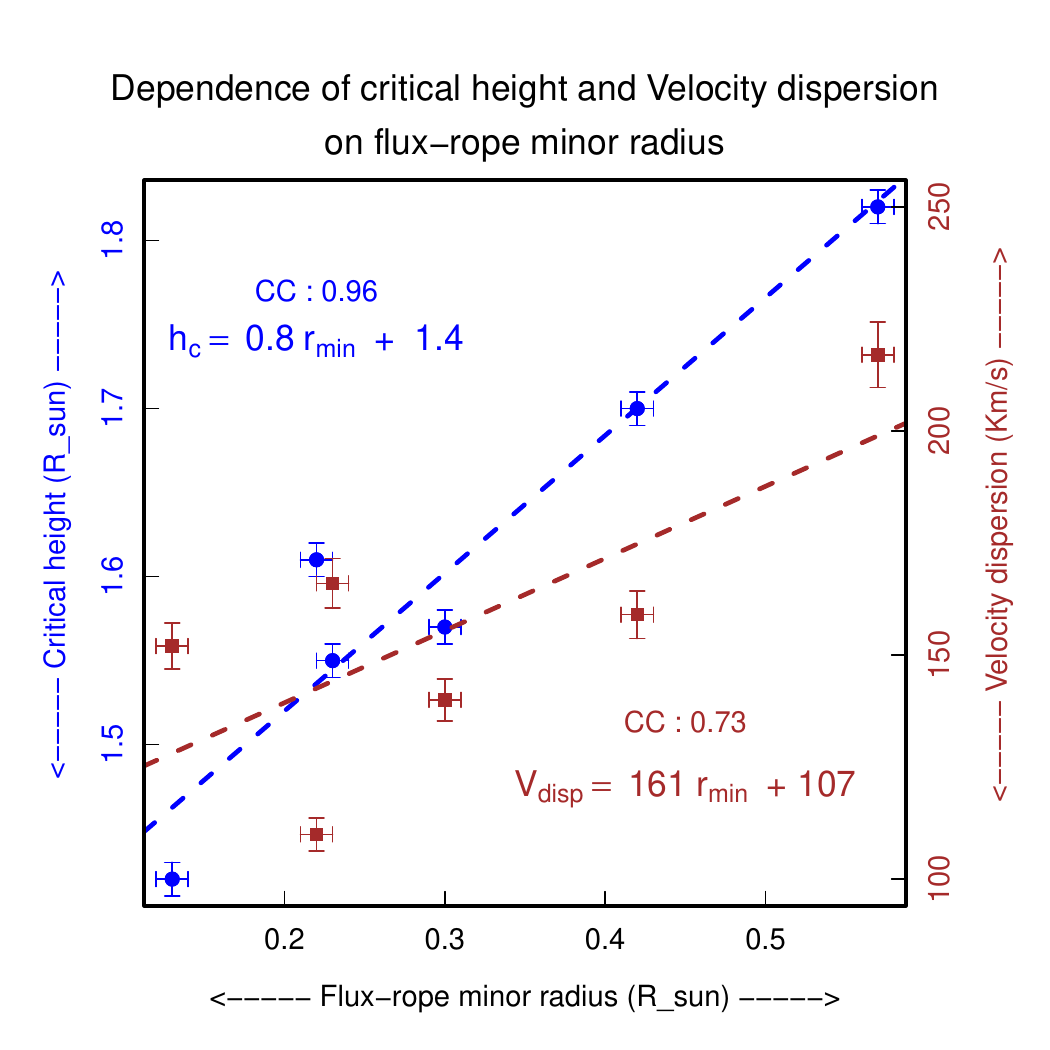}{0.48\textwidth}{(a)}
          \fig{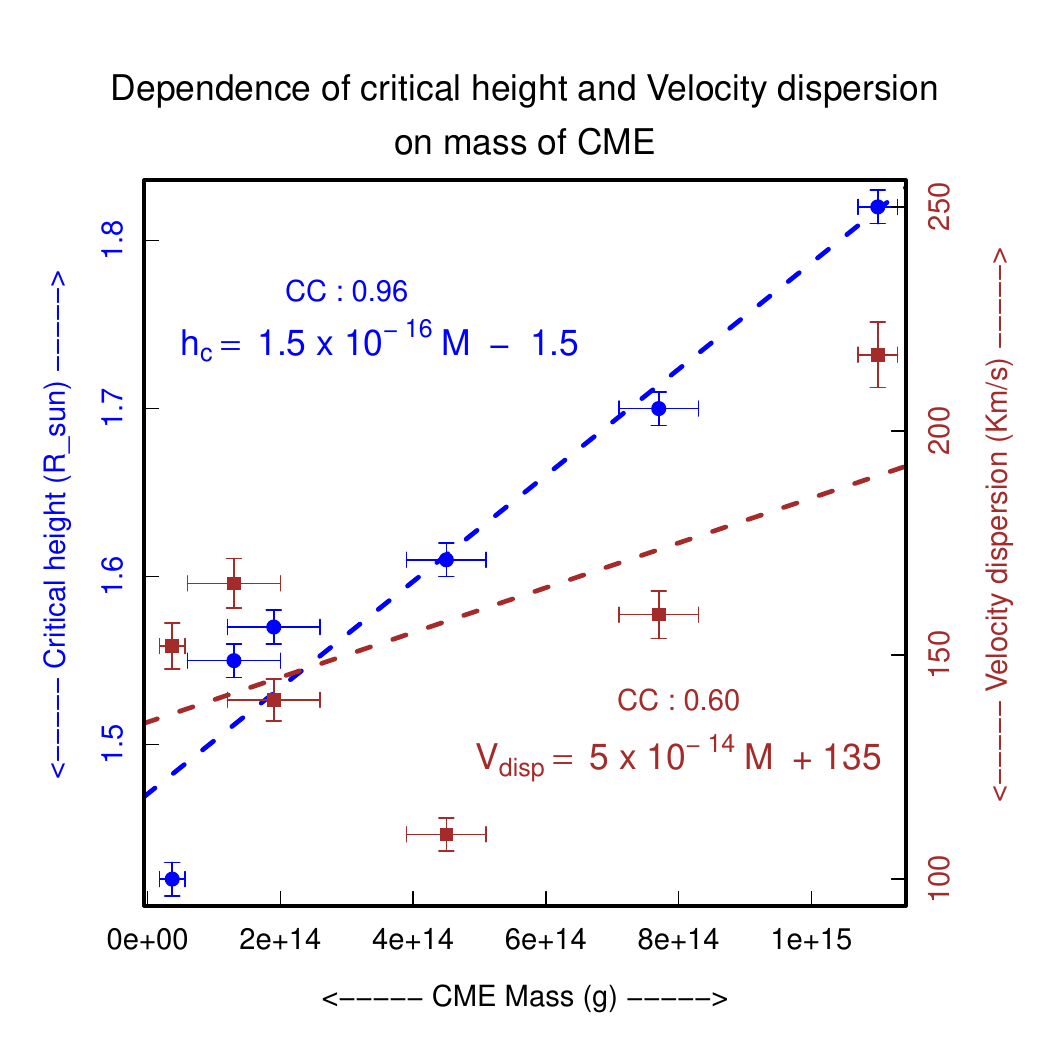}{0.48\textwidth}{(b)}
          }
\caption{A plot of the critical height (left-side scale) and magnitude of velocity dispersion (right-side scale) for all the CMEs, versus (a) flux-rope minor radius, (b) CME mass.
\label{prop}}
\end{figure*}

The phenomenon of velocity dispersion is also influenced by certain geometrical and morphological properties of CMEs. In Figure~\ref{prop}(a) we plot the variation of critical height (along the left hand vertical axis) and the magnitude of velocity dispersion, calculated by taking the difference between the average speeds of the LE and core, (along the right hand vertical axis) for all the CMEs, with respect to the CME flux-rope minor radius. The latter is calculated by fitting a circle to the flux-rope cross section (as shown in Figure~\ref{cme}(a)) and the corresponding radius of curvature is noted. The circle was fitted to the CME assuming a flux-rope morphology of the observed CME, where the flux-rope is embedded in the CME cavity, and a three part structure with the bright LE and core indicates a rough estimate of the extent of the flux-rope minor radius of curvature along the CME cross-section \citep[see ][for reference]{gopalswamy_2011}. We find that for the CMEs studied in this work, the critical height is higher for CMEs with larger flux-rope minor radius, with a strong positive correlation of $0.96$. A regression line is also fitted which relates the two quantities through the following equation:

\begin{equation}
    \mathrm{h_c} \, = \, 0.8\mathrm{r_{min}} \, + \, 1.4
\end{equation}

where $\mathrm{h_c}$ and $\mathrm{r_{min}}$ are in R$_{\odot}$. Thus a CME with a larger flux-rope cross-section is expected to experience the onset of dispersion at a higher height. We also find a positive correlation of the magnitude of dispersion with the flux-rope minor radius, with a correlation coefficient of $0.73$, indicating that once the dispersion sets in, the larger the flux-rope cross-section, the more the dispersion in velocity. This is expected, since the magnetic tension has an inverse dependence on the radius of curvature, and hence, the greater the flux-rope minor radius, the lower is the magnetic tension experienced, leading to greater dispersion in the velocities.

\subsection{Dependence on CME Mass}

In Figure~\ref{prop}(b), we similarly look into the dependence of critical height and magnitude of velocity dispersion on the mass of the CME. The mass of the CME is calculated using the SCC\_CALC\_CME\_MASS.PRO procedure (which is a part of the STEREO Science Center Software\footnote{https://stereo-ssc.nascom.nasa.gov/software.shtml}) in the Solarsoft package of IDL, which further allows the user to specify the angle of propagation of the CME with respect to the plane of the sky. In this regard, it should be noted that the mass of the CME is expected to increase, as the CME is known to accrete mass in the initial heights. Thus, we measured the mass in successive heights, and the final maximum mass was taken. We find that the critical height is strongly and positively correlated with the mass of the CME, with a correlation coefficient of $0.96$, with the two quantities related by the following equation,

\begin{equation}
    \mathrm{h_c}\,=\,1.5\,\times\,10^{-16}\mathrm{M}\,-\,1.5
\end{equation}

where $\mathrm{h_c}$ is in R$_{\odot}$ and M in g. Thus, from this we note that the more mass a CME has, the later is the onset of dispersion, or the more reluctant it is towards dispersion. Thus, it seems the velocity dispersion is an inertial property of the CME, as the higher the mass of a body, the greater the inertia, and hence the more reluctance towards exhibiting dispersion. We also find for these six CMEs, a positive correlation of $0.60$ between the dispersion magnitude and the mass of the CME, which further confirms the above idea, as once the dispersion sets in, the more massive CMEs are prone to more dispersion than the lighter ones.


\section{Discussion and conclusion} \label{conclusions}
In this work, using the combined observations from SWAP, K-Cor and LASCO-C2, the phenomenon of velocity dispersion inside CMEs is probed. Velocity dispersion has been studied in very few studies earlier and a crucial missing element was the implications of this dispersion on the kinematics in the inner coronal heights. For example, \cite{wood_1999,koutchmy_2008} study the relative kinematics of the different parts of a CME, but capture the late evolutionary phase beyond the inner coronal heights. Our analysis tries to address this particular missing element in our understanding of CMEs by studying the imprints of velocity dispersion on the kinematics in the inner corona. Prior to this work, there have been a few studies on velocity dispersion in the lower heights \citep[see ][]{schmahl_1977,nandita_2000,krall_2001,maricic_2004,chifu_2012}. However, although \cite{schmahl_1977} successfully captured the early evolution of the associated prominence, they failed to capture the evolution of the LE at the lower heights. This restricted them from comparing the relative evolution of the LE and core simultaneously in the lower heights. \cite{krall_2001} despite studying the velocity profiles in the lower heights for the different parts of the CMEs, were unable to arrive at clear conclusions due to challenges with the large noise in the data. The current work, with the help of the data from K-Cor, SWAP and LASCO-C2, successfully captures the height-time and velocity-time profiles for both the LE and the core in a combined field of view from 1.1 - 6 R$_{\odot}$, thus including the kinematics of CMEs happening in the inner corona as well as the middle corona \citep{west_2023}. Now, although the above studies focussed on the height-time profiles, our work extended their findings by probing the height-time evolution of the separation between the two structures, thereby finding the presence of a critical height ($\mathrm{h_c}$) which marks the onset of velocity dispersion. \cite{nandita_2000} studied a gradual CME and found that the acceleration of the core was delayed with respect to the LE, while \cite{maricic_2009} noted variations in the onset of acceleration of the two parts for a few CMEs, while the majority of the CMEs showed well synchronised accelerations of the LE and core. Now, whether the different critical heights reported in this work are an outcome of these variations in the acceleration phases can be explored in future studies. Our results show that the speeds of the core is less than the LE speeds for all the CMEs, which is also in agreement with the previous studies. Another important aspect of this study is the connection of the kinematics with the source region information. Although we do not find any clear imprint of the source regions on the estimated critical heights, yet we do find that the critical heights lie relatively closer to the Sun for CMEs coming from ARs, but the study needs to be extended to a larger sample set of CMEs to arrive at a strong conclusion on the effect of source regions.  A set of six CMEs were analysed in this work. In the following, we briefly summarize our main results:

\begin{itemize}
    \item A clear evidence of velocity dispersion through a radial velocity gradient is noted inside CMEs from the inner core to the LE in the inner corona. We report for the first time on the existence of a certain critical height $\mathrm{h_c}$ that marks the onset of velocity dispersion inside CMEs (Figure~\ref{h_diff}(b)). This is further supported by the results from a breakout CME simulated using MPI-AMRVAC, where a critical height is observed as well.

    \item A look into the speed profiles of the LE and the core in K-Cor for two CMEs provided two distinct pictures. For the impulsive CME (Figure~\ref{vel_prof}(a)), the LE speed profile shows the presence of impulsive acceleration phase, while the core does show a similar profile, but to a lesser extent. On the other hand, for the gradual CME (Figure~\ref{vel_prof}(b)), we find that both the LE and the core experience a small gradual acceleration which is reflected into a steady rise in the speeds. 
    \item We find that the critical height shows a strong positive correlation with the CME minor radius (Figure~\ref{prop}(a)), indicating that the larger the minor radius, the more the CME resists dispersion, and hence the higher the critical height. The magnitude of dispersion is also positively correlated with the minor radius. Thus, it seems a larger minor radius of CME resists dispersion for a longer time, but once the dispersion sets in, the strength of the dispersion is higher for larger minor radius CMEs.
    \item A strong positive correlation is also found between the mass of the CMEs and the critical height (Figure~\ref{prop}(b)), indicating the fact that the more massive a CME is, the later the dispersion starts. Furthermore, a positive correlation between the mass of the CME and the dispersion strength shows that the more massive the CME, the greater is the dispersion experienced once the CME crosses the critical height. Thus from these two factors, it is clear that velocity dispersion is an inertial property of a CME, as it seems that initially the mass resists the start of dispersion by increasing the critical height for more massive CMEs, but once the critical height is reached, the more massive CMEs experience more severe dispersion in the speeds of their LE and core. 
\end{itemize}

Thus, it is clear from the above results that the early kinematics of CMEs lead to a velocity dispersion inside CMEs, that leaves imprints on different kinematic properties of CMEs. However, it should be noted that the above conclusions are based on six events, hence a more elaborate statistical study will be important to have a better understanding of this phenomenon. Also, since this work uses single vantage point observations of CMEs (which introduces projection effects), in future, a study of velocity dispersion in the 3D kinematic profiles (found from 3D CME reconstruction techniques) would be a crucial follow up work along these lines. Also, recent and upcoming solar missions like Solar Orbiter \citep{muller_2020}, the recently launched Aditya-L1 \citep{prasad_2017} and the upcoming PROBA-3 \citep{shestov_2021} missions (will) have coronagraphs that (will) observe the inner corona with overlapping fields of view and with varied and improved image cadence \citep[for a comparison, see; ][]{nitin_2023}. These observations could be employed to study velocity dispersion in 3D by using different reconstruction techniques. We believe this work will improve our understanding of the early kinematics of CMEs in the lower heights, and would motivate more studies along these lines in the near future. Apart from that, these results will also provide crucial inputs to models that study CME initiation.
\begin{acknowledgments}
We thank the anonymous reviewer for the insightful comments and suggestions which have greatly helped in improving the manuscript. SM and DB acknowledge the PROBA-2 Guest Investigator program grant and the Royal Observatory of Belgium for all the cooperation throughout this project. SM acknowledges Vaibhav pant and Ritesh Patel for the very fruitful discussions regarding this work. The authors acknowledge NCAR for making the K-Cor (DOI: 10.5065/D69G5JV8) data available. This research was funded in whole, or in part by the Austrian Science Fund (FWF)[P34437]. SWAP is a project of the Centre Spatial de Li\`ege and the Royal Observatory of Belgium funded by the Belgian Federal Science Policy Office (BELSPO). Courtesy of the Mauna Loa Solar Observatory, operated by the High Altitude Observatory as part of the National Center for Atmospheric Research (NCAR). NCAR is supported by the National Science Foundation. The SOHO/LASCO data used here are produced by a consortium of the Naval Research Laboratory (USA), Max-Planck-Institut fuer Aeronomie (Germany), Laboratoire d’Astronomie (France), and the University of Birmingham (UK). SOHO is a project of international cooperation between ESA and NASA. The authors would like to acknowledge the support of Aryabhatta Research Institute of Observational Sciences (ARIES) and their Computer Section for facilitation and access to the ``Surya" High-Performance Computing (HPC) Facility in the completion of this work. The ROB team thanks the Belgian Federal Science Policy Office (BELSPO) for the provision of financial support in the framework of the PRODEX Programme of the European Space Agency (ESA) under contract numbers 4000120800, 4000134474, and 4000136681. The authors acknowledge the benefits from the discussion of the ISSI-BJ Team ``Solar eruptions: preparing for the next generation multi-wavelength coronagraph".
\end{acknowledgments}

%

\vspace{5mm}
\facilities{PROBA-2/SWAP, MLSO/K-Cor, SOHO/LASCO-C2}


\software{IDL, R.
          }



\appendix
\section{Supplementary Images}
\subsection{SWAP and K-Cor Images}
\begin{figure*}[ht!]
\gridline{\fig{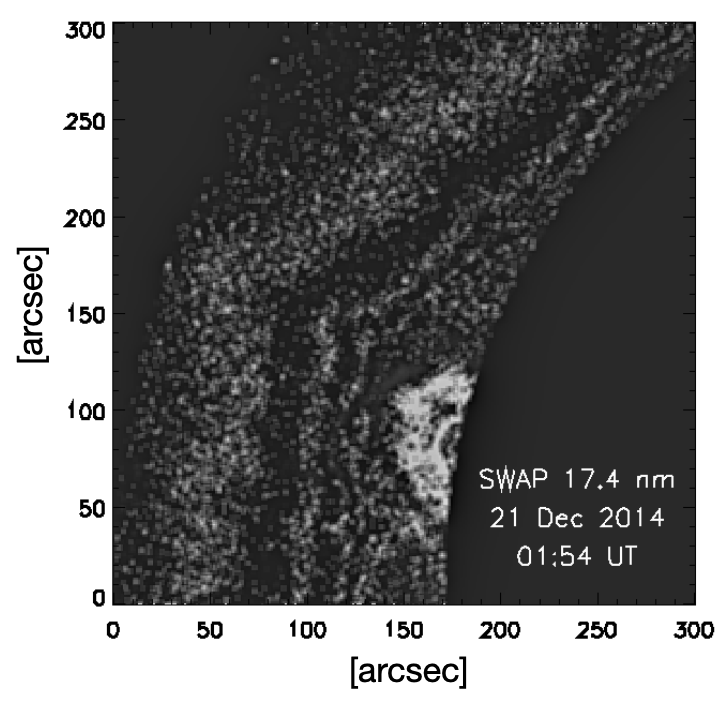}{0.3\textwidth}{(a)}
          \fig{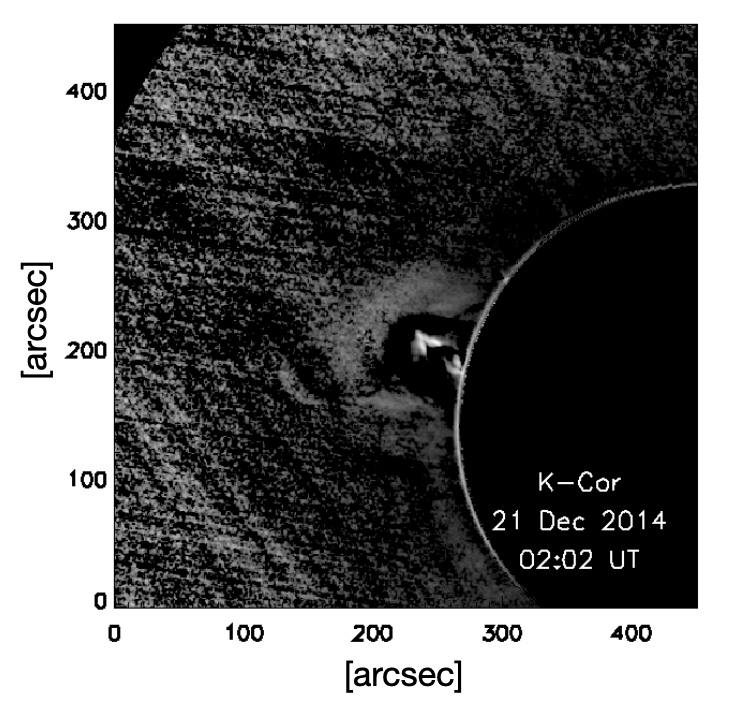}{0.3\textwidth}{(b)}
          }
\gridline{
          \fig{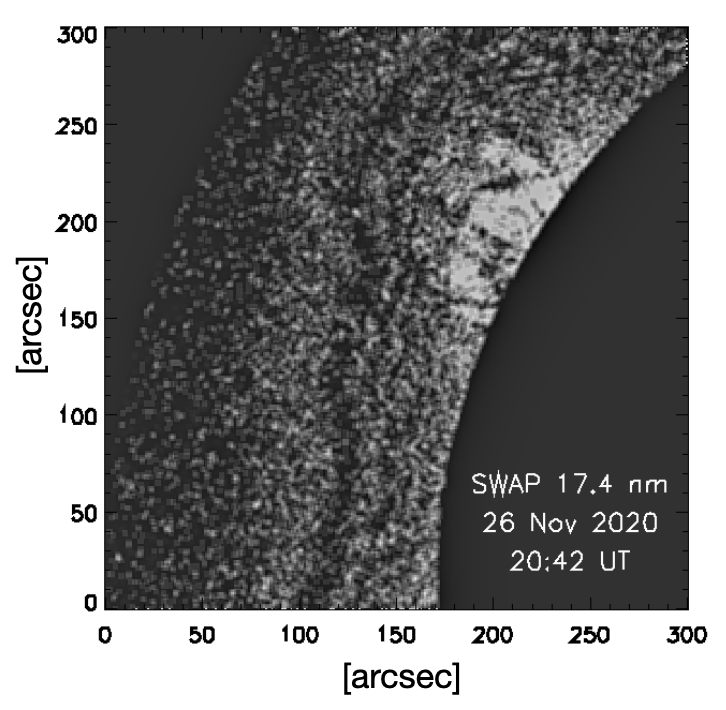}{0.3\textwidth}{(c)}
          \fig{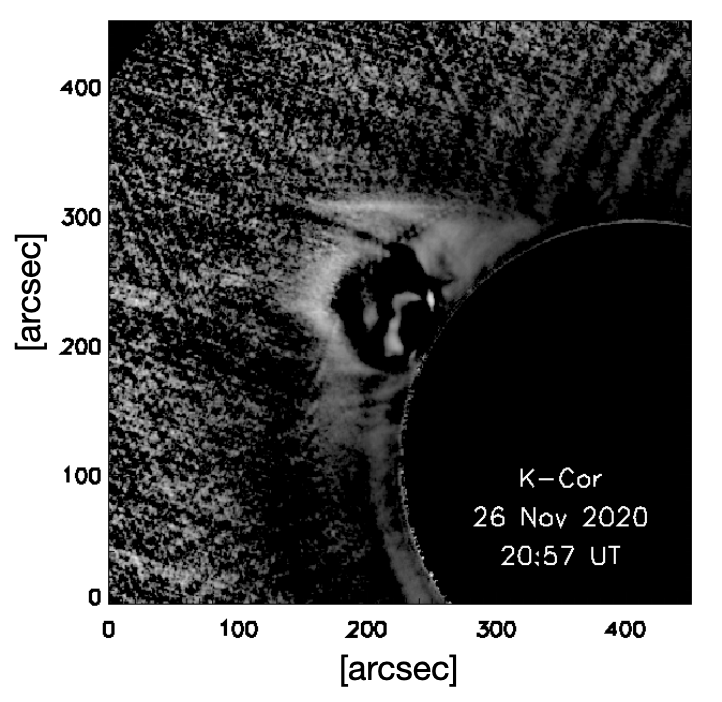}{0.3\textwidth}{(d)}
          }
\gridline{
          \fig{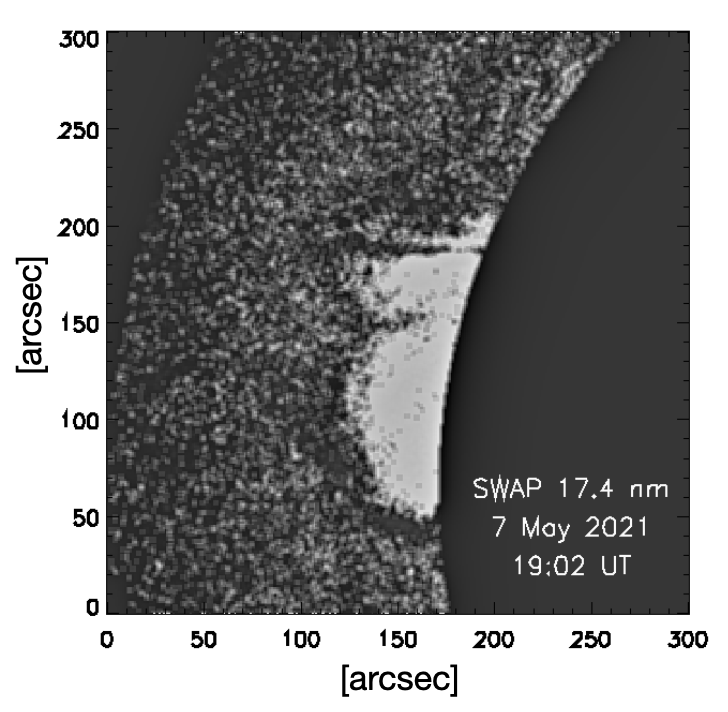}{0.3\textwidth}{(e)}
          \fig{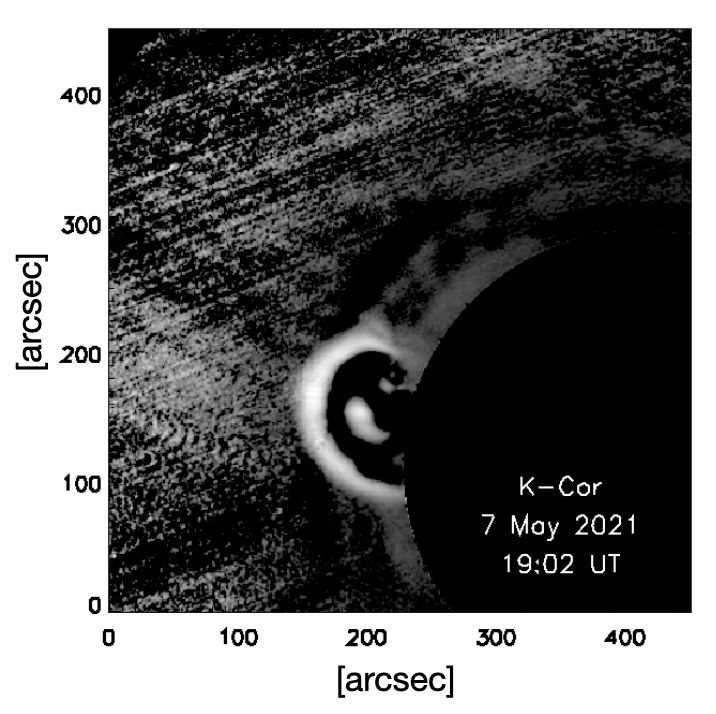}{0.3\textwidth}{(f)}
          }
\caption{Snapshots of the CMEs occurring on 21 December 2014 (top row), 26 November 2020 (middle row) and 7 May 2021 (bottom row) for the eruption front as observed by SWAP (left column) and for the three part structured CMEs as seen in the K-Cor (right column) field of view.
\label{append_fig1}}
\end{figure*}

\begin{figure*}[ht!]
\gridline{\fig{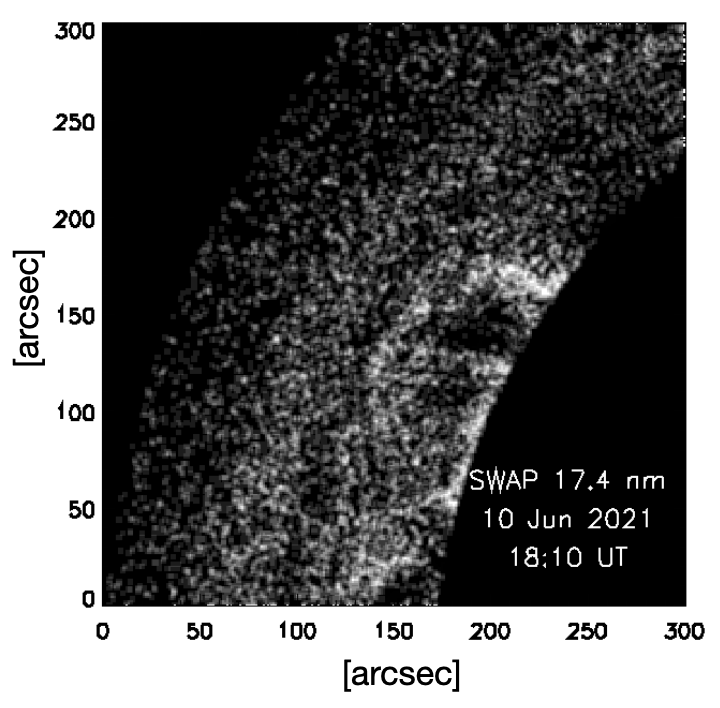}{0.3\textwidth}{(a)}
          \fig{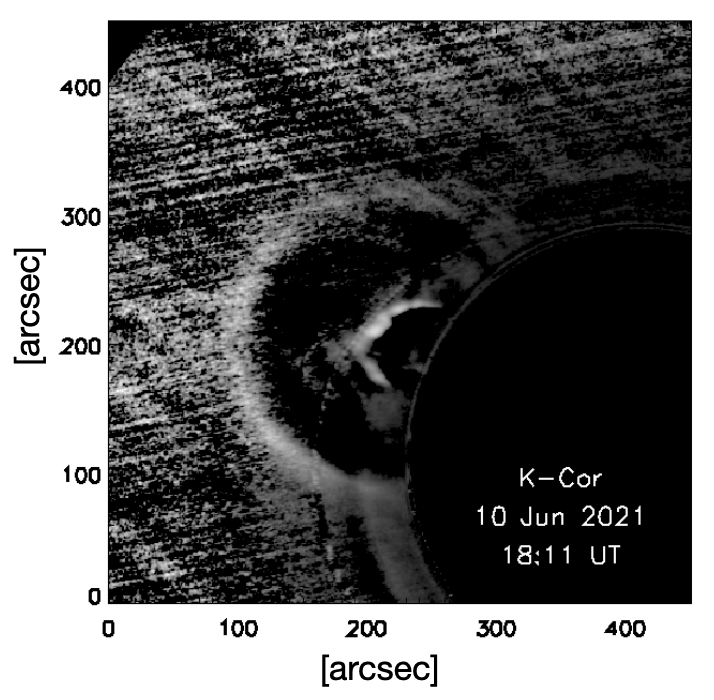}{0.3\textwidth}{(b)}
          }
\gridline{
          \fig{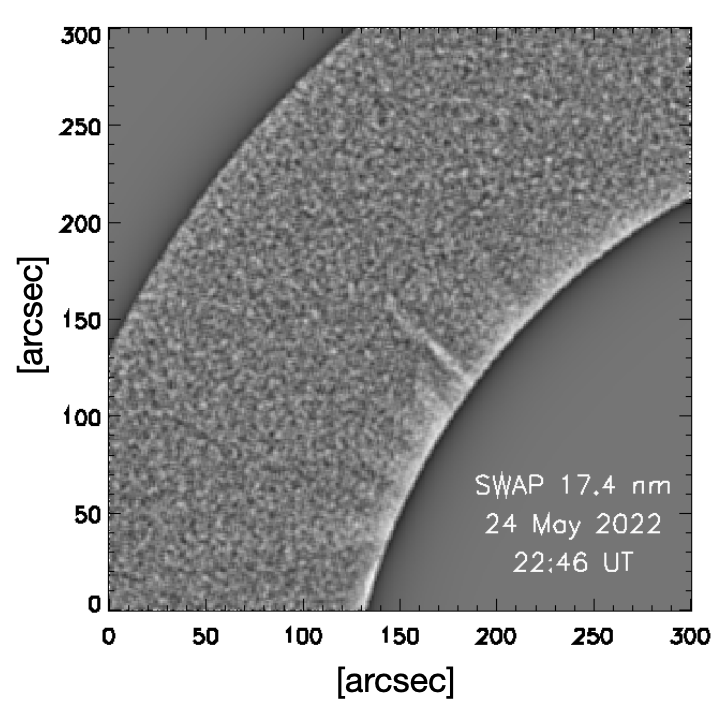}{0.3\textwidth}{(c)}
          \fig{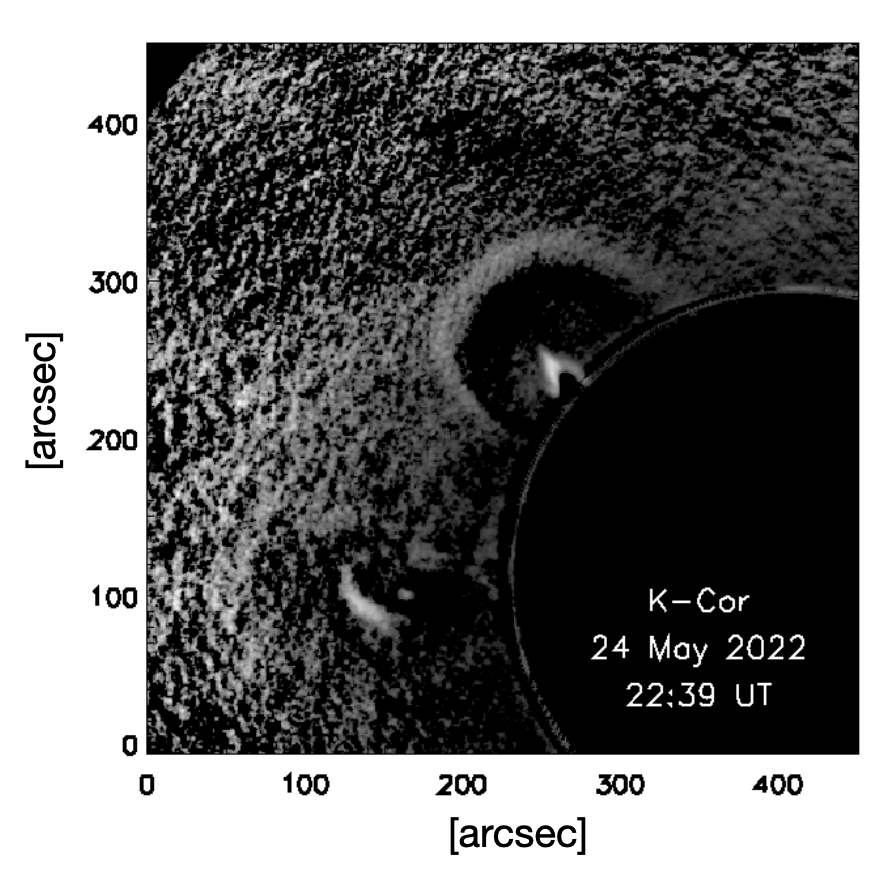}{0.3\textwidth}{(d)}
          }
\caption{Snapshots of the CMEs occurring on 10 June 2021 (top row), and 24 May 2022 (bottom row) for the eruption front as observed by SWAP (left column) and for the three part structured CMEs as observed in K-Cor (right column) field of view.
\label{append_fig2}}
\end{figure*}

In Figure~\ref{append_fig1} and \ref{append_fig2}, the corresponding SWAP and K-Cor images for the other five CMEs out of the six CMEs studied in this work are shown. Figure~\ref{append_fig1} shows the eruption front for the CME on 21 December 2014 in SWAP FOV in (a) and the three part structured CME in K-Cor FOV in (b). Similarly, for the CME on 26 November 2020, the eruption front in SWAP is shown in (c) and the CME in K-Cor in (d), and for the CME on 7 May 2021, the SWAP observation is shown in (e) and the K-Cor image in (f). Figure~\ref{append_fig2} shows the eruption front for the CME on 10 June 2021 in SWAP FOV in (a) and the CME in (b), and for the CME on 24 May 2022, the eruption front in SWAP is shown in (c) and the CME in K-Cor in (d).

\subsection{Kinematic Profiles}
\begin{figure*}[ht!]
\gridline{\fig{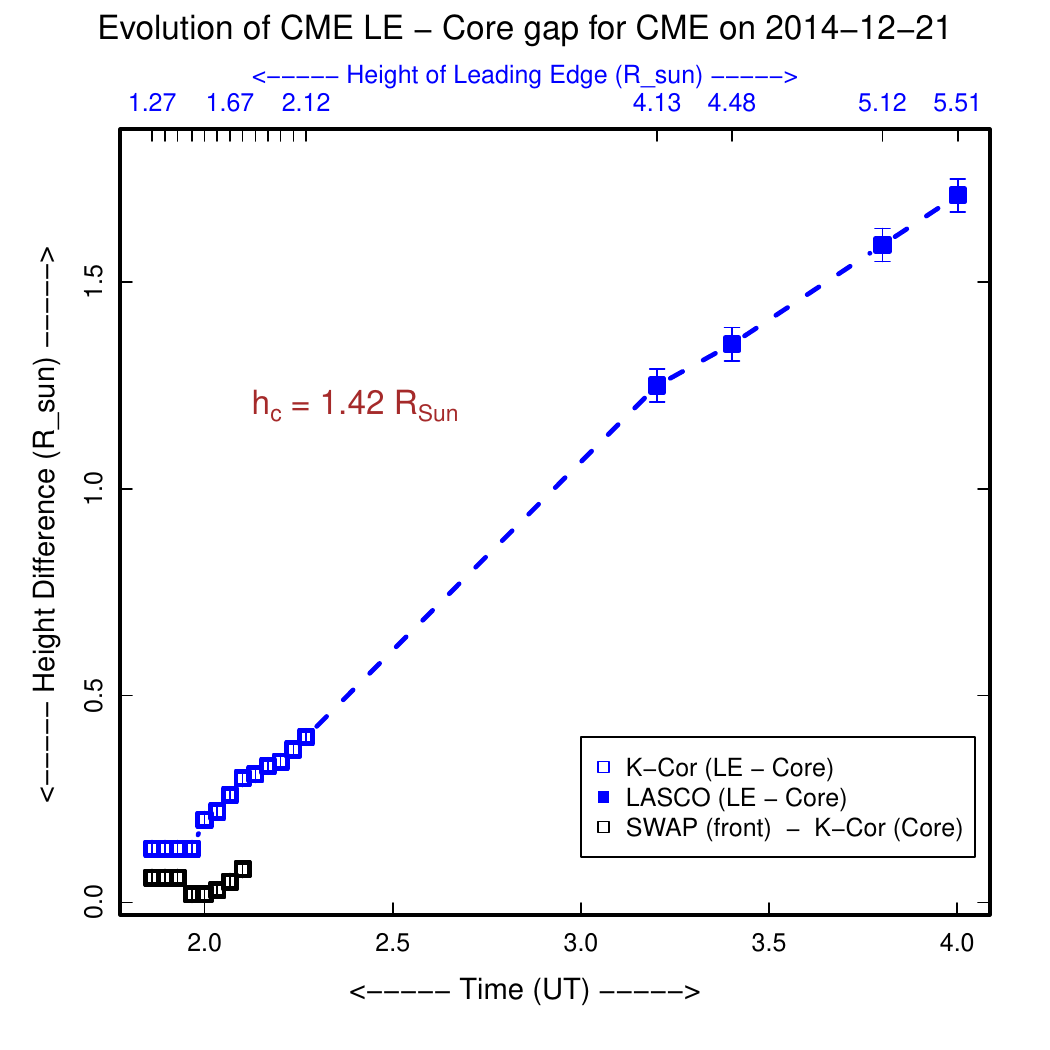}{0.33\textwidth}{(a)}
          \fig{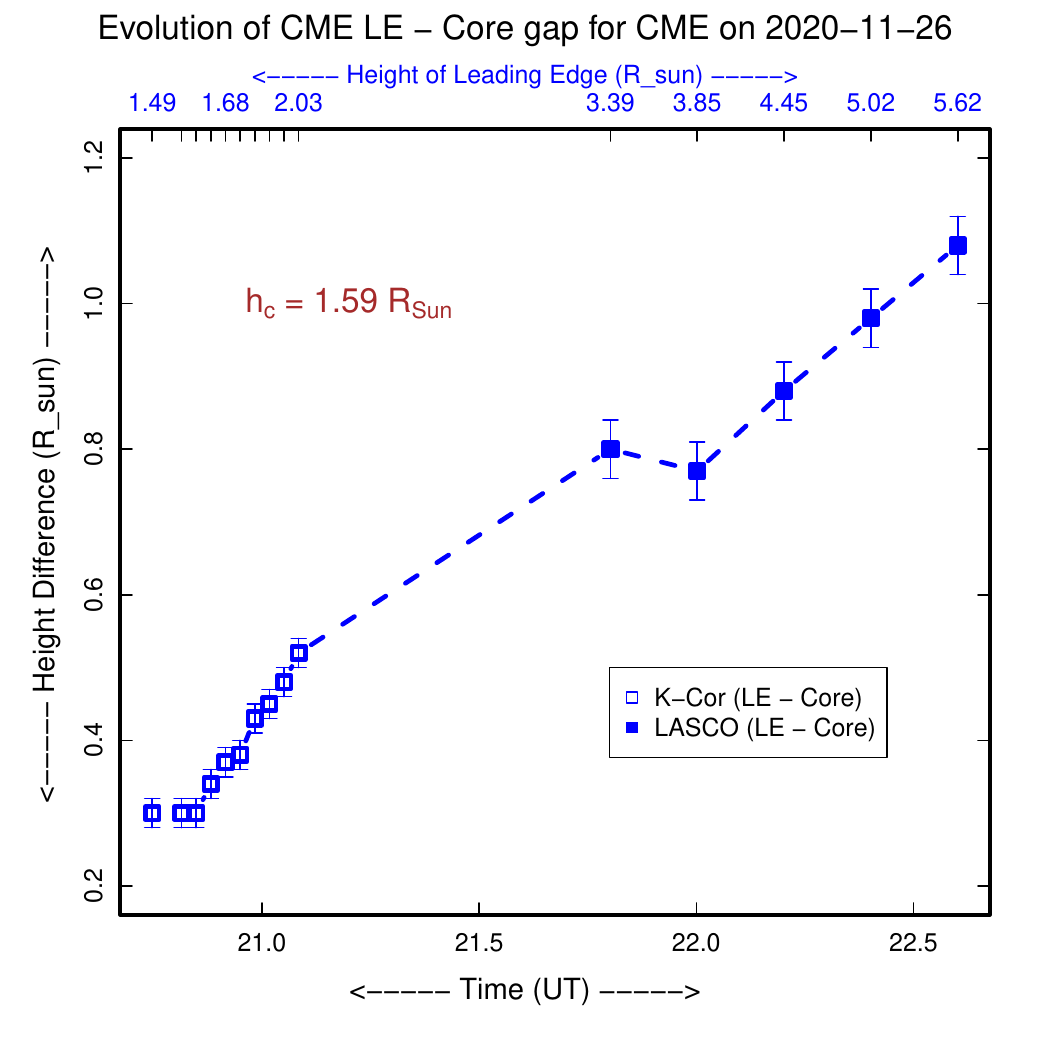}{0.33\textwidth}{(b)}
          }
\gridline{
          \fig{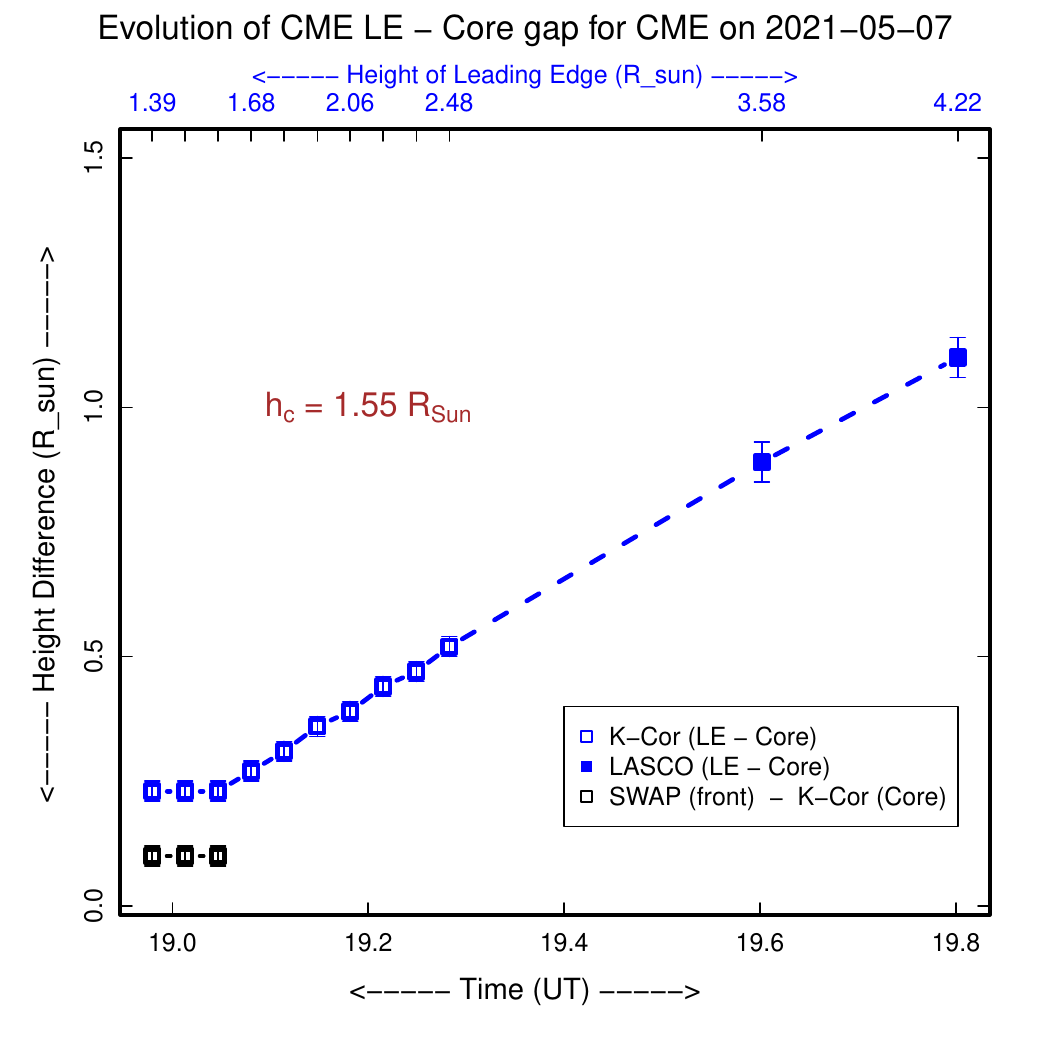}{0.33\textwidth}{(c)}
          \fig{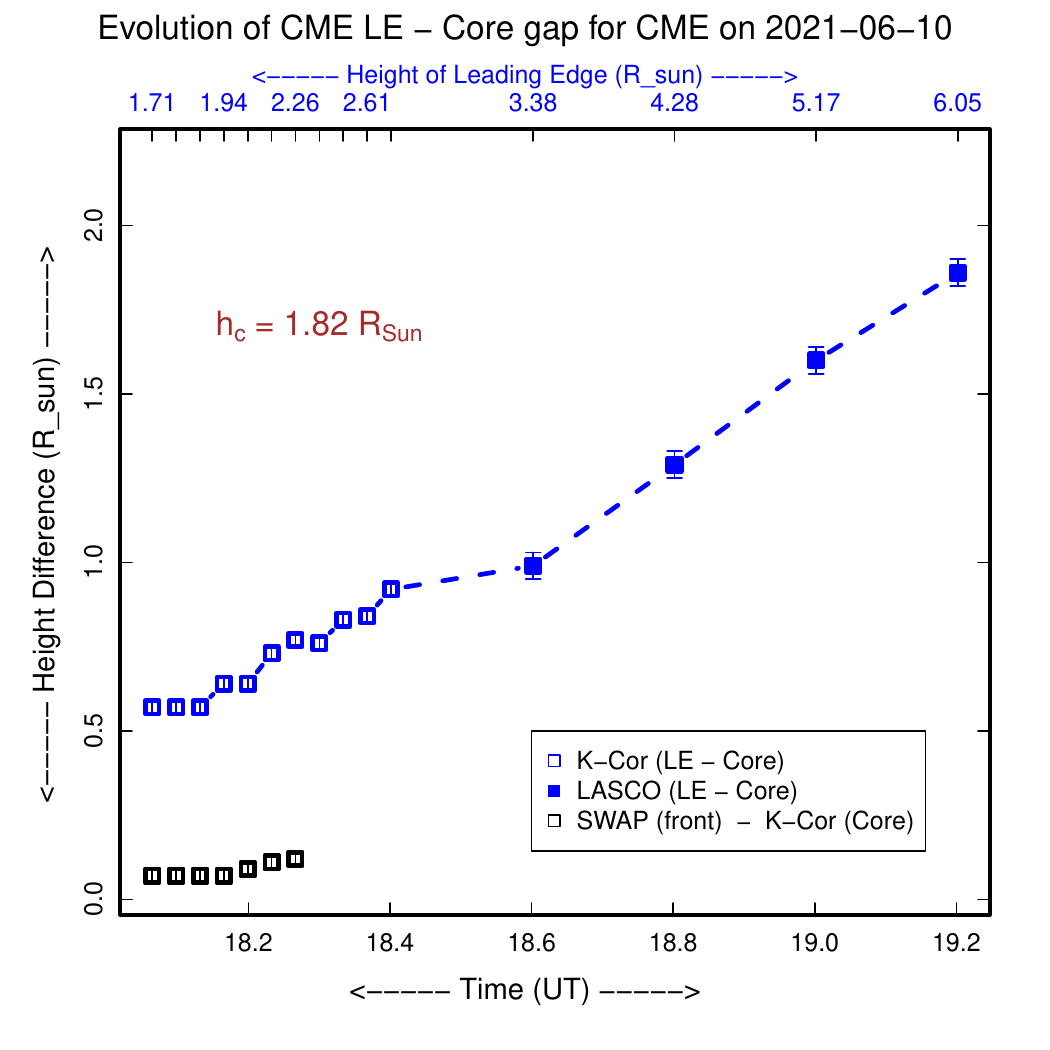}{0.33\textwidth}{(d)}
          }
\gridline{
        \fig{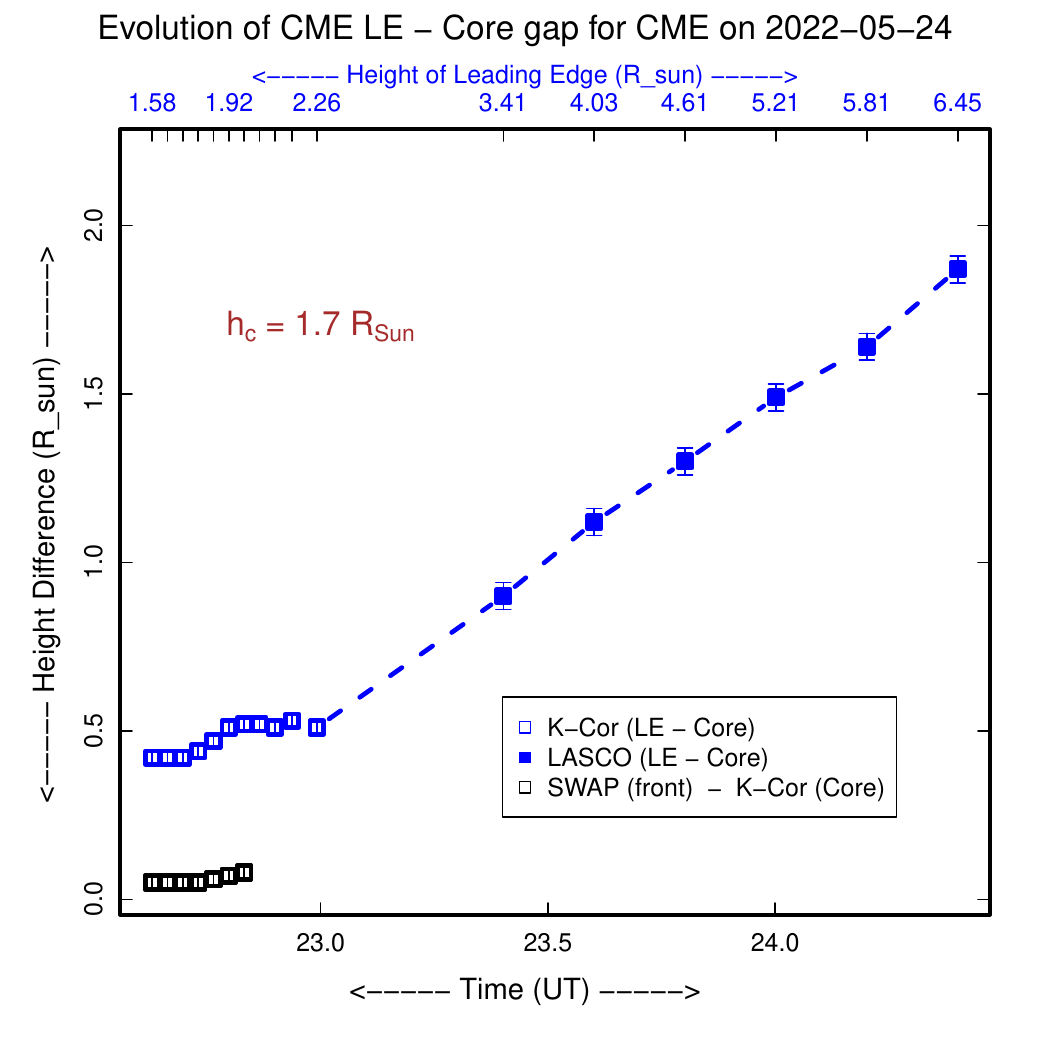}{0.33\textwidth}{(e)}
}
\caption{Plot of the difference in the height of the LE and core in the combined K-Cor and LASCO FOV, and the difference between SWAP eruption front and K-Cor core in the SWAP FOV for the CMEs  on 21 December, 2014 (a), 26 November, 2020 (b), 7 May, 2021 (c), 10 June, 2021 (d) and 24 May, 2022 (e). Please note that in (b), the corresponding SWAP observations were not available for this particular CME.
\label{append_fig4}}
\end{figure*}

\begin{figure*}[ht!]
\gridline{\fig{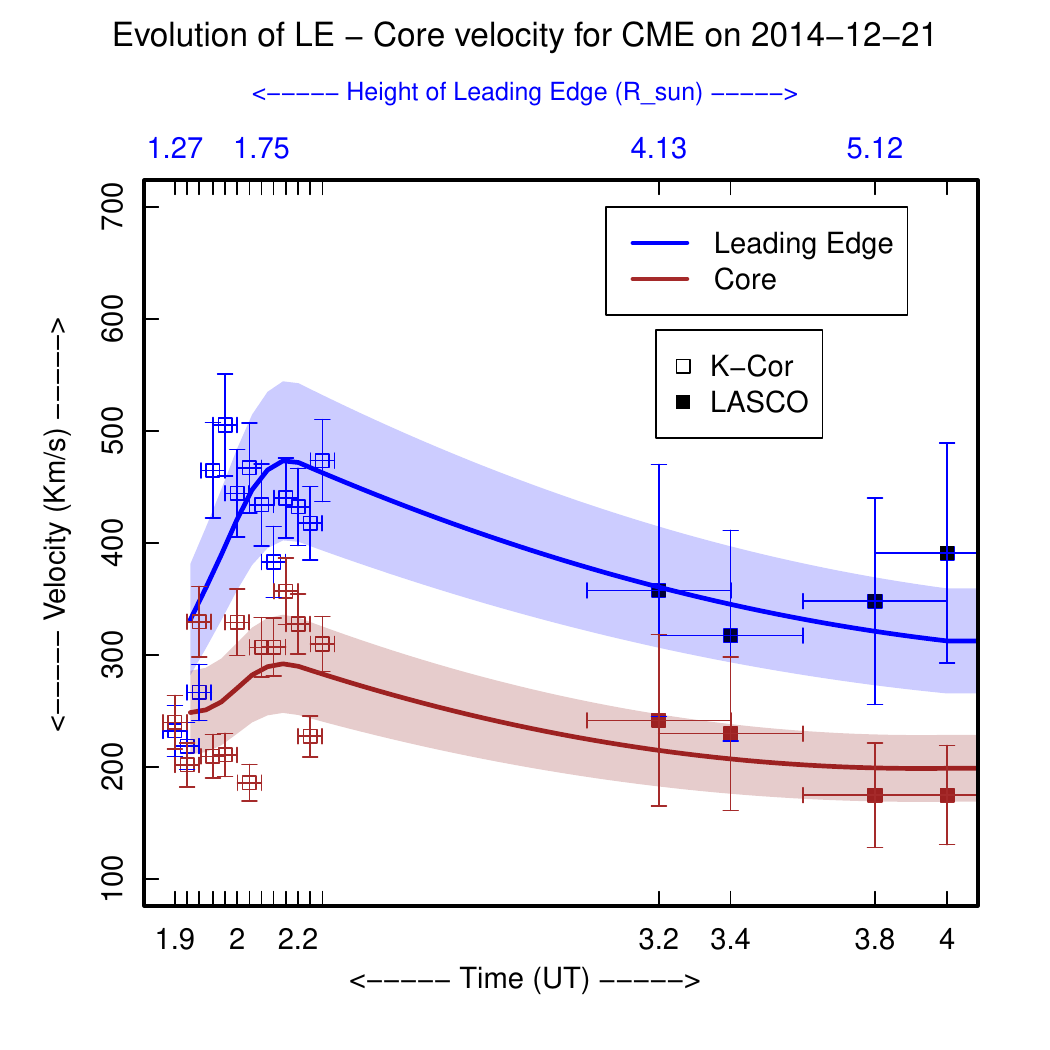}{0.48\textwidth}{(a)}
          \fig{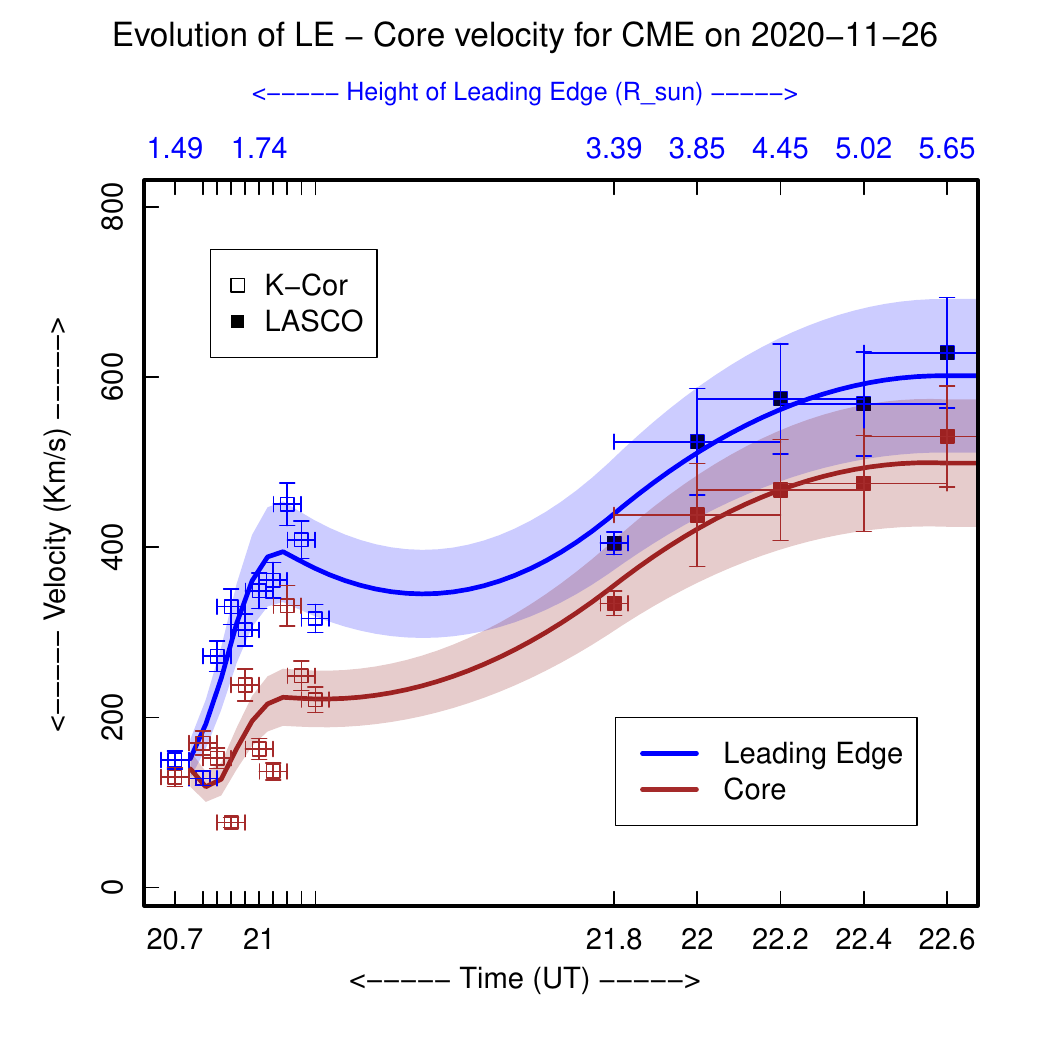}{0.48\textwidth}{(b)}
          }
\gridline{
          \fig{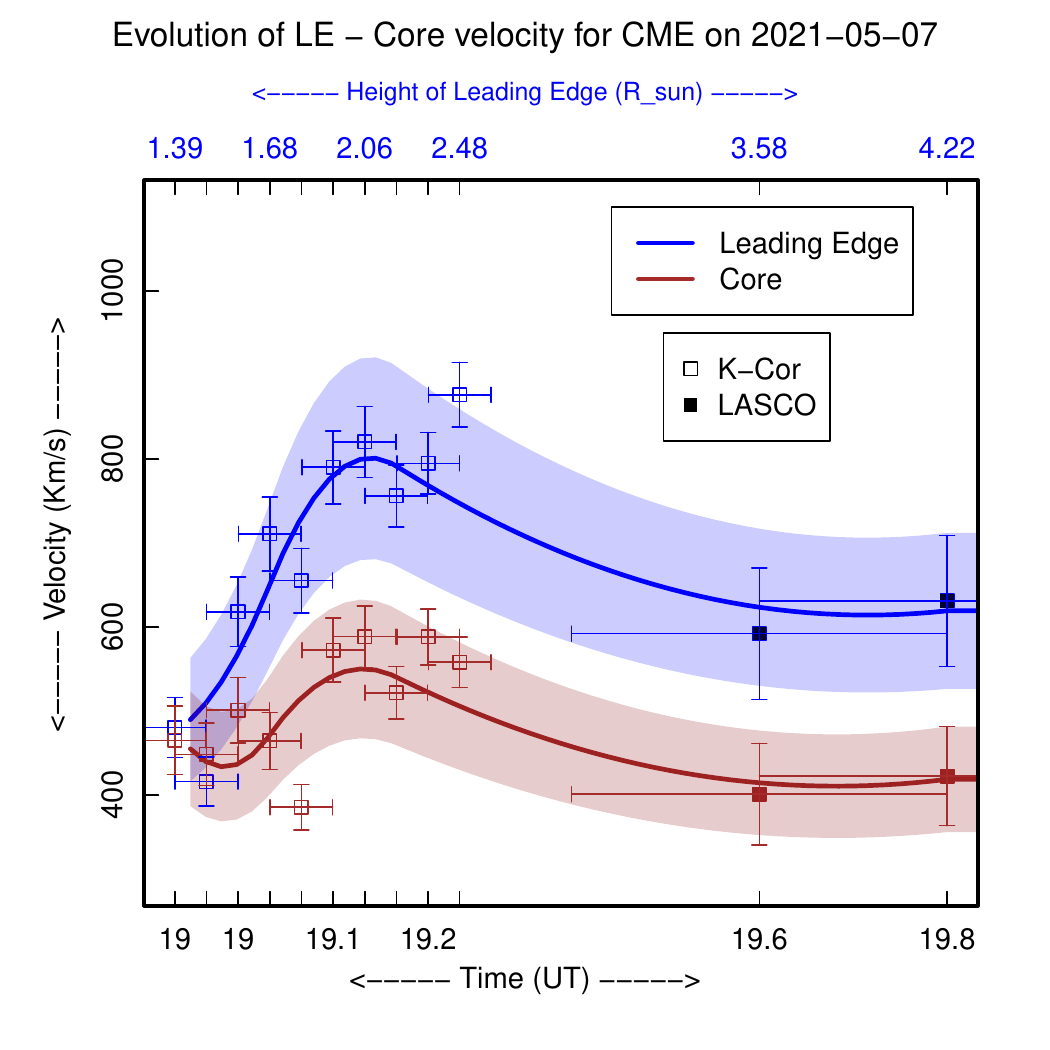}{0.48\textwidth}{(c)}
          \fig{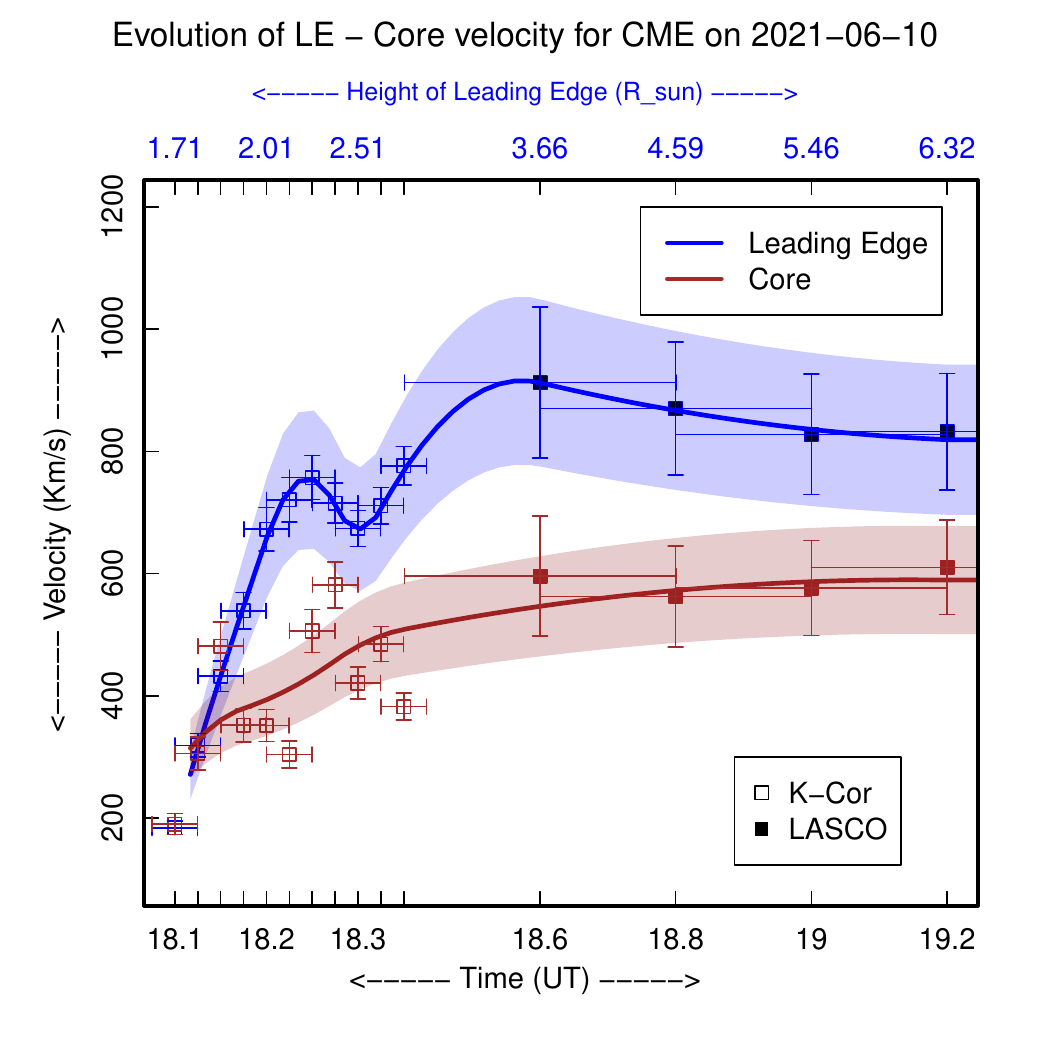}{0.48\textwidth}{(d)}
          }
\caption{Velocity profiles of the LE and core in the combined K-Cor and LASCO FOV for the CMEs on 21 December, 2014 (a), 26 November, 2020 (b), 7 May, 2021 (c) and 10 June, 2021 (d).
\label{append_fig3}}
\end{figure*}

In Figure~\ref{append_fig3}, the velocity profiles for the CME LE and core for the other four CMEs studied in this work are shown. It can be seen again that although both the LE and core experience acceleration, but the magnitude of the experienced accelerations are different, as the LE is clearly seen to travel with much higher speeds than the core. The evolution of the difference in heights between the LE and core in K-Cor and LASCO images and the difference in heights between the eruption front in SWAP and core in K-Cor are shown in Figure~\ref{append_fig4} for the other five CMEs studied in this work. It can be seen from Figure~\ref{append_fig4}, that the evolution of the height difference indicates the height of start of velocity dispersion for all the events. Please note that the evolution of the gap between the eruption front in SWAP and the K-Cor core is not shown for the CME on 26 November 2020, as for that CME, the corresponding SWAP observations were not available. It is worth pointing out that with better image cadence in the inner coronal observations, it would be possible in future studies by combining data from Aditya-L1, PROBA-3, to better capture and understand the initial phase before the start of velocity dispersion \citep[for a comparison of the different image cadence of different upcoming missions, see][]{nitin_2023}.  

\end{document}